\documentclass[lettersize,journal]{IEEEtran}
\usepackage{amsmath,amsfonts,amsthm}
\usepackage{algorithmic}
\usepackage{algorithm}
\usepackage{array}
\usepackage[caption=false,font=normalsize,labelfont=sf,textfont=sf]{subfig}
\usepackage[hyphens]{url}
\usepackage{graphicx}
\usepackage[numbers,sort&compress]{natbib}
\usepackage{caption}
\usepackage{booktabs}
\usepackage{amsmath}
\usepackage{amsfonts}
\usepackage{bm}
\usepackage[table]{xcolor}
\usepackage{multirow}

\definecolor{bestgreen}{HTML}{B9FBC0}
\newcommand{\best}[1]{\cellcolor{bestgreen}\textbf{#1}}
\newcommand{\second}[1]{\underline{#1}}

\newcommand{\method}{WaveOp-LiteFM}
\newcommand{\sevir}{SEVIR}
\newcommand{\cfm}{conditional flow matching}
\definecolor{orange}{RGB}{247, 224, 213}

\begin{document}

\title{WaveOp-LiteFM: Lightweight Neural-Operator Flow Matching for Satellite-to-Radar Precipitation Retrieval}

\author{Chunlei~Shi, Yecheng~Zhang, Yufeng~Zhu, Dan~Niu, Yichao~Dong, Yongchao~Feng, and Junming~Hou

\thanks{This work was supported by the  Heavy Rainfall Research Foundation of China (No. BYKJ2025M14), China Meteorological Administration Xiong\textrm{'}an Atmospheric Boundary Layer Key Laboratory (No. 2025LABL-B12), and by the National Natural Science Foundation of China (62374031, 62331009), and by NSFC-Jiangsu Province (BK20240173). (\textit{Corresponding author: Dan Niu, Yongchao Feng and Junming Hou.})}
\thanks{Chunlei Shi, Yufeng Zhu, Yichao Dong and Dan Niu are with the Department of Automation, Southeast University, Nanjing 210096, China (e-mail: danniu1@163.com).}% <-this % stops a space
\thanks{ 
Yecheng Zhang is with the Department of Architecture, Tsinghua University, Beijing 100084, China. 
Yongchao Feng is with the State Key Laboratory of Virtual Reality Technology and Systems, Beihang University, Beijing 100191, China (e-mail: 230238514@seu.edu.cn).
Junming Hou is with the State Key Laboratory of Millimeter Waves, School of Information Science and Engineering, Southeast University, Nanjing 210096, China (e-mail: junming\_hou@seu.edu.cn).}}

\markboth{Journal of \LaTeX\ Class Files,~Vol.~14, No.~8, August~2021}%
{Shell \MakeLowercase{\textit{et al.}}: A Sample Article Using IEEEtran.cls for IEEE Journals}


\maketitle

\begin{abstract}
% Satellite-based radar retrieval is widely used to complement ground-based radar observations in regions affected by terrain blockage, limited detection range, or incomplete radar coverage.
% Satellite-to-radar (S2R) retrieval refers to estimate ground radar precipitation from geostationary satellite observations, providing a crucial solution for precipitation monitoring in radar-sparse regions affected by terrain blockage, limited detection range, or incomplete radar coverage. 
Satellite-to-radar (S2R) retrieval refers to estimating ground-based radar precipitation from geostationary satellite observations, enabling precipitation monitoring in regions with limited radar coverage.
While recent generative flow matching models have greatly advanced retrieval quality, they face a critical trade-off: pixel-space formulations suffer from prohibitive computational costs of attention-based U-Net velocity networks, whereas latent-space modeling often sacrifices fine precipitation details or struggles with sparse targets.
To break this dilemma, we propose \method{}, a lightweight neural operator flow matching framework for S2R retrieval. 
Our approach introduces a novel velocity backbone built upon the fundamental spectral-local-wavelet (SLW) block, enabling efficient and stable modeling of flow matching in pixel-space. Specifically, the SLW block disentangles precipitation features into three distinct frequency regimes: (i) the spectral branch captures large-scale stratiform organization;(ii) the local branch models short-range interactions; and (iii) the Wavelet branch enhances sharp structures while suppressing noisy high-frequency responses.
Building on this, an input-adaptive gating mechanism dynamically fuses the features from three functional branches. Furthermore, a skip gate efficiently reintegrates encoder features via additive fusion within the decoder, circumventing the costly channel concatenation in conventional U-Net architectures. 
Experiments on \sevir{} and southeast China datasets show that \method{} achieves state-of-the-art retrieval while substantially reducing computational costs.
% only 2.61M parameters and about $6.7\times$ lower sampling FLOPs.
Beyond benchmark evaluation, large-area inference over China, such as the recent Typhoon Bavi case, demonstrates that \method{} maintains reliable retrieval quality in large-scale real-world scenarios.
% To address these limitations, we propose \method{}, a lightweight neural-operator flow matching framework for satellite-to-radar precipitation retrieval.
% \method{} keeps the stable pixel-space conditional flow matching objective, but replaces the heavy velocity backbone with a spectral-local-wavelet (SLW) block that separates low-frequency precipitation organization from high-frequency boundary and detail modeling.
% Specifically, the spectral branch captures storm-scale intensity organization, the depthwise local branch models short-range convective interactions, and the Haar wavelet shrinkage branch emphasizes sharp meteorological boundaries and localized radar details while suppressing unstable high-frequency responses.
% Building on this frequency-aware decomposition, an input-dependent branch gate dynamically weights the three operators for each feature state rather than using a fixed fusion rule.
% Furthermore, a skip gate injects encoder details through additive decoder fusion, avoiding the expensive high-channel concatenation used in conventional U-Net velocity networks.
% Experiments on the public \sevir{} and southeast China datasets show that \method{} achieves state-of-the-art retrieval performance with only 2.61M parameters and about $6.7\times$ lower sampling FLOPs.
% Beyond benchmark evaluation, large-area inference over China, including a recent Typhoon Bavi case, further suggests that \method{} maintains reliable retrieval quality at broader spatial scales while substantially reducing the cost of generative satellite precipitation retrieval.
\end{abstract}

% \begin{IEEEkeywords}
% Satellite-to-radar retrieval, Lightweight Neural-Operator, Flow Matching.
% \end{IEEEkeywords}

\section{Introduction}
Satellite-based radar retrieval is crucial for synthesizing radar-like precipitation products for short-range warning, flood monitoring, aviation weather, and convective-storm analysis \citep{han2026llm}.
However, ground radar networks often suffer from limited spatial coverage, terrain blockage, range effects, heterogeneous quality control, and operational gaps \citep{an2026toward,si2023novel}.
Geostationary satellites provide frequent wide-area observations from visible, infrared, and lightning-related channels, making a practical solution to extend radar-like precipitation monitoring beyond radar coverage.
Despite this operational value, satellite-to-radar retrieval has received less attention than radar extrapolation and general precipitation nowcasting \citep{shi2026langretrieval}.
Furthermore, real-world deployment demands models that can efficiently process large-scale regional grids under operational or edge-computing constraints while preserving fine-scale convective cores and sharp precipitation boundaries.

\begin{figure*}[htbp]
  \centering
  \includegraphics[width=\linewidth]{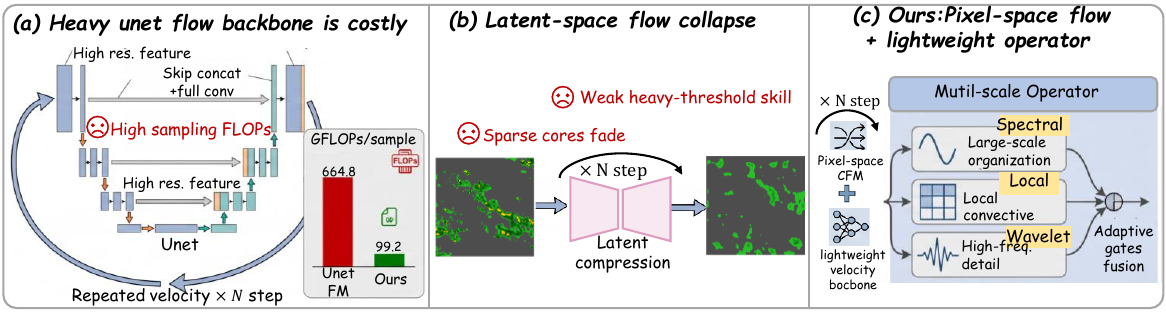}
  % \caption{
  % Motivation of lightweight satellite-to-radar precipitation retrieval.
  % Satellite observations provide wide-area cloud and convection cues, while radar-like precipitation fields require localized intensity and structure.
  % The key challenge is to keep the stable pixel-space generative supervision needed for sparse precipitation targets while reducing the velocity-network cost for regional and large-area deployment.
  % }
  % \caption{
  % Comparison of satellite-to-radar retrieval paradigms.
  % Satellite inputs provide wide-area cloud and convection cues, whereas radar-like targets require localized precipitation intensity and structure.
  % Compared with latent-space or heavy pixel-space generative retrieval, \method{} preserves direct pixel-space supervision while reducing velocity-network cost for regional and large-area inference.
  % }
  \caption{
  Comparison of generative satellite-to-radar retrieval paradigms.
  (a) Heavy pixel-space U-Net flow is accurate but costly; (b) latent-space flow is efficient but may smooth sparse precipitation cores; and (c) \method{} keeps pixel-space supervision with lightweight spectral-local-wavelet operators.
  }
  \label{fig:motivation}
  % \vspace{-15pt}
\end{figure*}

Deep neural networks have become common tools for precipitation nowcasting, satellite-radar fusion, and weather retrieval \citep{qin2026srdiff,wu2026mocast,ravuri2021skilful,espeholt2022deep,zhang2023skilful}.
Deterministic convolutional or transformer-based models provide efficient direct mappings, but they often produce overly smooth precipitation fields and weaken localized convective cores.
Recent diffusion and flow-based models improve retrieval sharpness by learning generative precipitation distributions \citep{gao2023prediff,lipman2023flow,liu2022flow,schusterbauer2026probabilistic}.
However, in high-resolution regional deployment, pixel-space generative samplers repeatedly evaluate velocity or score networks, and conventional U-Net backbones become especially expensive because dense feature processing and skip concatenation are performed at every sampling step (Fig.~\ref{fig:motivation}a).

To alleviate the memory, storage, and computational burden of high-resolution weather modeling, recent studies have increasingly shifted weather forecasting, nowcasting, and generative modeling from pixel space to compact latent or compressed representations \citep{leinonen2023latent,zhao2025transforming,wang2026longwang}.
These methods reduce the spatial resolution of the modeled state and can lower repeated sampling or prediction cost.
For satellite-to-radar retrieval, however, the effectiveness of latent-space modeling depends critically on preserving spatially sparse and localized precipitation structures \citep{chen2022rainnet}.
In practice, we observed that a more aggressive latent operator-flow variant can become difficult to optimize for sparse precipitation fields, especially on \sevir{} VIL, where the model may prefer diffuse or background-dominated predictions.
Existing alternatives therefore leave an unresolved trade-off: pixel-space generative models preserve precipitation details but are computationally heavy, whereas aggressive latent transport improves efficiency at the risk of losing localized radar structure (Fig.~\ref{fig:motivation}b).

To address this challenge, we propose \method{}, a spectral-local-wavelet operator flow matching model.
Concretely, \method{} parameterizes the \cfm{} velocity field with a three-branch operator block.
A Fourier spectral branch models broad precipitation organization and long-range dependencies; a depthwise separable local branch captures convective cells and local boundaries; and a Haar wavelet threshold branch learns to shrink noisy high-frequency coefficients while preserving radar-like details.
Building on these complementary operators, adaptive gates select the useful mixture of branches for each feature map.
Meanwhile, gated additive skip fusion in the decoder avoids expensive skip concatenation.
The resulting fully convolutional architecture can be applied to different spatial sizes, while operational large-area inference can be handled through patch stitching (Fig.~\ref{fig:motivation}c). Our contributions are summarized as follows:

\begin{itemize}
  % \item We propose \method{}, a lightweight neural-operator flow matching framework for satellite-to-radar precipitation retrieval, which preserves stable pixel-space supervision while substantially reducing the cost of repeated generative sampling.
  \item We propose \method{}, a lightweight neural-operator flow matching framework for satellite-to-radar precipitation retrieval that enables high-fidelity pixel-space modeling while significantly improving computational efficiency.
  % \item We introduce a spectral-local-wavelet (SLW) block that couples spectral, depthwise local, and Haar wavelet shrinkage operators to capture storm-scale organization, short-range convective interactions, and sharp radar-like boundary details.
 \item We introduce a principled spectral-local-wavelet (SLW) block that disentangles precipitation representations into complementary frequency regimes, with adaptive dual-gating mechanisms enabling dynamic feature fusion and efficient skip connections.
  % \item We design adaptive gated fusion mechanisms, including input-dependent branch gating and gated additive skip fusion, to dynamically combine complementary operators and preserve encoder details without expensive high-channel skip concatenation.
  \item Beyond achieving state-of-the-art performance on standard benchmarks, we further demonstrate the model’s scalability and robustness in large-scale real-world scenarios through tiled inference over the recent Typhoon Bavi in China.
\end{itemize}

\section{Related Work}

\paragraph{Precipitation nowcasting and satellite-radar retrieval.}
Deep learning has become a standard tool for precipitation nowcasting, from convolutional recurrent models to attention, transformer, diffusion, and loss-driven designs \citep{shi2015convolutional,shi2017deep,wang2017predrnn,bai2022rainformer,gao2022earthformer,gao2023prediff,yan2024fourier}.
Large-scale systems such as DGMR, MetNet, and NowcastNet further show that learned precipitation models can operate at practically relevant scales \citep{sonderby2020metnet,ravuri2021skilful,espeholt2022deep,zhang2023skilful}.
Efficiency-oriented nowcasting has also been studied via physically correlated channel grouping in lightweight U-Nets \citep{wang2026phygroup}.
Unlike radar extrapolation, satellite-to-radar retrieval must infer radar-like intensity and convective cores from indirect satellite evidence; we reduce the repeated velocity-network evaluations required by pixel-space generative retrieval.

\begin{figure*}[t]
  \centering
  \includegraphics[width=\linewidth]{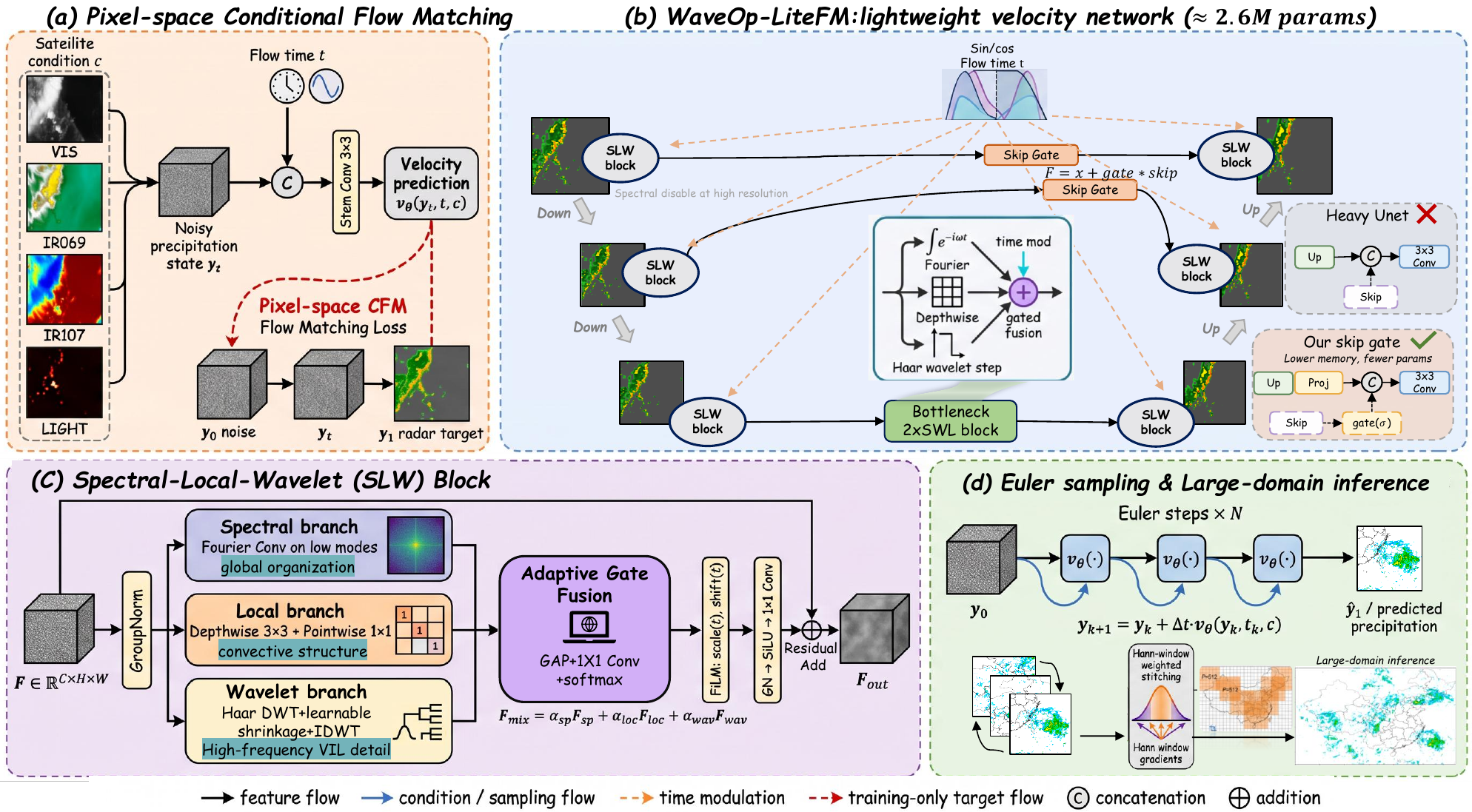}
  % \caption{
  % Technical framework of \method{}.
  % The model keeps pixel-space conditional flow matching for direct radar-field supervision and replaces the heavy U-Net velocity backbone with a lightweight spectral-local-wavelet operator network.
  % % Spectral, local, and wavelet branches capture complementary precipitation structures and are fused through adaptive gates, while gated skip fusion reduces decoder computation.
  % }
  \caption{
  % Overview of the \method{} architecture for satellite-to-radar retrieval.
  Overview of the \method{}.
  Pixel-space conditional flow matching provides direct radar-field supervision, while spectral-local-wavelet operators replace the heavy U-Net velocity backbone and use adaptive branch and skip gates to capture complementary precipitation structures with reduced decoder computation.
  }
  % \vspace{-15pt}
  \label{fig:framework}
\end{figure*}

\paragraph{Generative and latent-space weather modeling.}
Diffusion, flow matching, and rectified flow models learn generative precipitation distributions through denoising or neural velocity fields, with flow matching offering a stable regression-style objective and efficient ODE sampling \citep{gao2023prediff,lipman2023flow,liu2022flow,schusterbauer2026probabilistic}.
Latent diffusion, cascaded latent models, weather autoencoders, and latent generative priors reduce high-resolution generation cost by operating on compact states before decoding \citep{leinonen2023latent,gong2024cascast,li2025extreme,zhao2025transforming,wang2026longwang}.
We instead keep conditional flow matching in pixel space for direct supervision of sparse precipitation, and reduce cost through the velocity-network design.

\paragraph{Operator and frequency-domain precipitation modeling.}
Neural operators learn mappings between functions and have been widely used in Fourier or adaptive Fourier forms for partial differential equations and data-driven weather forecasting \citep{lifourier,guibas2021adaptive,pathak2022fourcastnet}.
Frequency-domain objectives and decompositions are also effective for precipitation because average pixel losses often smooth intense, localized echoes \citep{yan2024fourier,lin2025alphapre}.
Motivated by these observations, \method{} integrates spectral, local, and wavelet operators into a conditional radar-retrieval velocity network, enabling multi-scale precipitation modeling within pixel-space flow matching.

\section{Method}
\label{sec:method}

\subsection{Problem Formulation}

For each retrieval case, let $c\in\mathbb{R}^{C_c\times H\times W}$ denote the satellite observation and $y\in\mathbb{R}^{C_y\times H\times W}$ denote the radar-like precipitation target.
In the southeast China dataset, $c$ contains two infrared channels and $y$ is a reflectivity-like field.
In \sevir{}, $c$ contains visible, IR069, IR107, and lightning channels, and $y$ is VIL.
The task is to learn a conditional generative mapping
\begin{equation}
  \hat{y} \sim p_\theta(y\mid c),
\end{equation}
that converts indirect cloud and convection signatures into a radar-like field.
This mapping has two conflicting requirements.
First, it must retain direct supervision on sparse high-intensity precipitation, since small displacements or over-smoothing can strongly affect threshold skill.
Second, it must remain efficient when the same sampler is applied many times or deployed on large regional grids.

\subsection{Framework Overview}

\method{} separates the pixel-space generative objective from the velocity-network parameterization.
Fig.~\ref{fig:framework} gives the overall architecture.
The generative process remains in pixel space for direct precipitation supervision, while the heavy U-Net velocity backbone is replaced by a lightweight operator-style encoder-decoder.
Given a noisy precipitation state $y_t$, a flow time $t$, and a satellite condition $c$, the network predicts
\begin{equation}
  v_\theta(y_t,t,c)
  =
  D_\theta
  \left(
  E_\theta([y_t,c]), \phi(t)
  \right),
  \label{eq:velocity-overview}
\end{equation}
where $E_\theta$ and $D_\theta$ denote the compact encoder and decoder, $[\cdot,\cdot]$ is channel concatenation, and $\phi(t)$ is the flow-time embedding.
The encoder-decoder is built from spectral-local-wavelet (SLW) blocks that model storm-scale organization, local convective interaction, and boundary-sensitive details.

\subsection{Pixel-Space Conditional Flow Matching}

To avoid latent-space degeneration on sparse precipitation targets, \method{} performs conditional flow matching directly in pixel space.
This keeps the transport objective tied to radar intensity and location before introducing the lightweight velocity parameterization.
We use a rectified-flow-style \cfm{} objective.
For a target radar field $y_1$ and Gaussian noise $y_0\sim\mathcal{N}(0,I)$, we sample $t\in(0,1)$ and construct the linear probability path
\begin{equation}
  y_t = (1-t)y_0 + t y_1 .
\end{equation}
where $y_t$ is the interpolated precipitation state between the noise sample $y_0$ and target radar field $y_1$.
The corresponding conditional velocity is
\begin{equation}
  u_t = y_1-y_0 .
\end{equation}
where $u_t$ gives the target transport direction from noise to radar observation.
The satellite-conditioned velocity network is optimized by
\begin{equation}
  \mathcal{L}_{\mathrm{FM}}
  =
  \mathbb{E}_{y_0,y_1,t}
  \left[
  \left\|
  v_\theta(y_t,t,c)-u_t
  \right\|_2^2
  \right].
  \label{eq:fm-loss}
\end{equation}
where $v_\theta(y_t,t,c)$ is the satellite-conditioned velocity prediction.
At inference, sampling starts from $y_0\sim\mathcal{N}(0,I)$ and uses Euler integration,
\begin{equation}
  y_{k+1}=y_k+\Delta t\,v_\theta(y_k,t_k,c)
  \label{eq:euler-sampling}
\end{equation}
where $t_k=k/K$, $\Delta t=1/K$, and $K$ is the number of sampling steps.
We keep the main transport in pixel space to preserve direct supervision on precipitation intensity and location.

\subsection{Spectral-Local-Wavelet Operator Block}

The SLW block is designed to retain the stability of pixel-space flow matching while reducing the cost of high-resolution velocity evaluation.
Instead of using high-channel U-Net convolutions, it couples three lightweight and complementary operators for global organization, local convective interaction, and high-frequency boundary detail.
Given $\mathbf{F}\in\mathbb{R}^{C\times H'\times W'}$, an SLW block can be written as
\begin{equation}
  \begin{aligned}
  \mathrm{SLW}(\mathbf{F},t)
  &=
  \mathbf{F}
  +
  h_\theta
  \left(
  \Gamma_t[\mathbf{F}_{\mathrm{mix}}]
  \right),\\
  \mathbf{F}_{\mathrm{mix}}
  &=
  \sum_{b\in\{\mathrm{sp},\mathrm{loc},\mathrm{wav}\}}
  \alpha_b(\mathbf{F})\,\mathcal{O}_{b,\theta}(\mathbf{F}).
  \end{aligned}
  \label{eq:slw-overview}
\end{equation}
where $\mathcal{O}_{b,\theta}$ denotes the spectral, local, or wavelet operator, and $\alpha_b(\mathbf{F})$ is its input-dependent branch weight.
$\Gamma_t$ denotes FiLM-style flow-time modulation, and $h_\theta$ is the output projection before the residual update.

\paragraph{Spectral branch.}
To capture broad precipitation organization, we design a Fourier spectral branch.
This choice is motivated by frequency-domain precipitation modeling, where low-frequency spectral components encode large-scale intensity layout and organized storm structure \citep{lin2025alphapre,an2026toward}.
The branch first compresses channels with a $1\times1$ bottleneck, applies Fourier convolution on retained low-frequency modes, and projects the result back to the block width:
\begin{equation}
  \widehat{\mathbf{Z}}_o(\mathbf{k})
  =
  \sum_i \mathbf{R}_{\theta,io}(\mathbf{k})
  \widehat{\mathbf{X}}_i(\mathbf{k}),
  \quad
  \mathbf{k}\in\mathcal{K},
\end{equation}
where $\widehat{\mathbf{X}}=\mathcal{F}(P_{\mathrm{in}}\mathbf{F})$, $\mathbf{F}_{\mathrm{sp}}=P_{\mathrm{out}}\mathcal{F}^{-1}(\widehat{\mathbf{Z}})$, $\mathcal{K}$ is the retained low-frequency mode set, and $\mathbf{R}_\theta$ contains learned complex mode-wise channel-mixing weights.
It is enabled from lower-resolution encoder levels, where global spectral mixing is more cost-effective.

\paragraph{Local branch.}
To preserve convective-scale locality, we use a depthwise separable local branch.
It provides short-range spatial mixing for compact cells and boundary adjustments without the cost of dense convolutions:
\begin{equation}
  \mathbf{F}_{\mathrm{loc}} = P_{1\times1}(D_{3\times3}(\mathbf{F})).
\end{equation}
where $D_{3\times3}$ is a depthwise convolution and $P_{1\times1}$ is a pointwise projection.

\paragraph{Wavelet branch.}
To emphasize sharp meteorological boundaries and localized radar texture, we design a Haar wavelet branch.
This follows the frequency-decoupling observation in WaveC2R that wavelet high-frequency components are sensitive to precipitation edges and fine-scale details \citep{shi2026wavec2r}.
The branch decomposes the feature map, shrinks unstable high-frequency responses, and reconstructs the result:
\begin{equation}
  (L,H_1,H_2,H_3)=\mathrm{DWT}(\mathbf{F}),
\end{equation}
\begin{equation}
  \tilde{H}_i=\mathrm{sign}(H_i)\max(|H_i|-\tau_\theta,0),
\end{equation}
\begin{equation}
  \mathbf{F}_{\mathrm{wav}}=\mathrm{IDWT}(L,\tilde{H}_1,\tilde{H}_2,\tilde{H}_3).
\end{equation}
where $L$ is the low-frequency Haar component, $H_i$ are the three high-frequency components, and $\tau_\theta$ is a learned channel-wise threshold.
This branch suppresses unstable high-frequency responses while retaining localized radar texture and sharp echo boundaries.

\paragraph{Adaptive branch fusion and time modulation.}
The three operators contribute differently across precipitation regimes: stratiform systems require more global organization, whereas compact convection relies more on local and wavelet details.
Instead of using a fixed fusion rule, the SLW block predicts an input-dependent branch gate:
\begin{equation}
  (\alpha_{\mathrm{sp}},\alpha_{\mathrm{loc}},\alpha_{\mathrm{wav}})
  =
  \mathrm{softmax}(g(\mathrm{GAP}(\mathbf{F}))),
\end{equation}
\begin{equation}
  \mathbf{F}_{\mathrm{mix}} =
  \alpha_{\mathrm{sp}}\mathbf{F}_{\mathrm{sp}}
  + \alpha_{\mathrm{loc}}\mathbf{F}_{\mathrm{loc}}
  + \alpha_{\mathrm{wav}}\mathbf{F}_{\mathrm{wav}}.
\end{equation}
where $\alpha_{\mathrm{sp}}$, $\alpha_{\mathrm{loc}}$, and $\alpha_{\mathrm{wav}}$ are softmax-normalized weights predicted from global pooled features.
After branch fusion, the flow time $t$ modulates the mixed feature,
\begin{equation}
  \tilde{\mathbf{F}}_{\mathrm{mix}}
  =
  \mathbf{F}_{\mathrm{mix}}\odot(1+s_\theta(\phi(t))) + b_\theta(\phi(t)),
\end{equation}
where $s_\theta(\phi(t))$ and $b_\theta(\phi(t))$ are feature-wise scale and shift terms generated from the flow-time embedding.
The final block output is residual:
\begin{equation}
  \mathbf{F}_{\mathrm{out}} = \mathbf{F} + h_\theta(\tilde{\mathbf{F}}_{\mathrm{mix}}).
\end{equation}

\begin{table*}[t]
  \centering
  \caption{
  Quantitative comparison with the controlled pixel-space flow baseline on SE China and \sevir{}.
  MAE$_s$ is reported on the dataset target scale.
  Heavy-threshold scores use 35 dBZ for SE China and encoded VIL@219 for \sevir{}; higher is better except for GFLOPs, MAE$_s$, and LPIPS.
  Improv. reports the relative gain of \method{} over LiteFM-UNet following each metric direction.
  The best score in each column is highlighted in \textcolor{green!60!black}{green}.
  }
  \label{tab:main_metrics}
  \small
  \setlength{\tabcolsep}{4pt}
  \renewcommand{\arraystretch}{1.08}
  \resizebox{\textwidth}{!}{
  \begin{tabular}{c l ccccccccc}
    \toprule
    Dataset & Method & Params $\downarrow$ & GFLOPs/sample $\downarrow$ & MAE$_s$ $\downarrow$ & PSNR $\uparrow$ & SSIM $\uparrow$ & LPIPS $\downarrow$ & CSI-heavy $\uparrow$ & POD-heavy $\uparrow$ & HSS-heavy $\uparrow$ \\
    \midrule
    % \multirow{6}{*}{\rotatebox{90}{SE China}} 
    \multirow{3}{*}{SE China}
    & LiteFM-UNet & 5.54M & 10116.6 
    & 3.010 & 19.58 & 0.555 & 0.368 & 0.094 & 0.122 & 0.168 \\
     & \method{} & \best{2.61M} & \best{1500.9} 
     & \best{2.725} & \best{20.31} & \best{0.578} & \best{0.343} & \best{0.146} & \best{0.197} & \best{0.250} \\
     & \cellcolor{gray!20}\textcolor{red}{\textit{Improv.}} & \cellcolor{gray!20}\textcolor{red}{+52.9\%} & \cellcolor{gray!20}\textcolor{red}{+85.2\%}
     & \cellcolor{gray!20}\textcolor{red}{+9.5\%} & \cellcolor{gray!20}\textcolor{red}{+3.7\%} & \cellcolor{gray!20}\textcolor{red}{+4.1\%} & \cellcolor{gray!20}\textcolor{red}{+6.8\%} & \cellcolor{gray!20}\textcolor{red}{+55.3\%} & \cellcolor{gray!20}\textcolor{red}{+61.5\%} & \cellcolor{gray!20}\textcolor{red}{+48.8\%} \\
    \midrule
    % \multirow{7}{*}{\rotatebox{90}{\sevir{}}} 
    \multirow{3}{*}{\sevir{}} 
    & LiteFM-UNet & 5.54M & 664.8 
    & 12.342 & 21.25 & 0.523 & 0.201 & 0.133 & 0.222 & 0.233 \\
     & \method{} & \best{2.61M} & \best{99.2} 
     & \best{11.985} & \best{21.32} & \best{0.548} & \best{0.192} & \best{0.151} & \best{0.289} & \best{0.260} \\
     & \cellcolor{gray!20}\textcolor{red}{\textit{Improv.}} & \cellcolor{gray!20}\textcolor{red}{+52.9\%} & \cellcolor{gray!20}\textcolor{red}{+85.1\%}
     & \cellcolor{gray!20}\textcolor{red}{+2.9\%} & \cellcolor{gray!20}\textcolor{red}{+0.3\%} & \cellcolor{gray!20}\textcolor{red}{+4.8\%} & \cellcolor{gray!20}\textcolor{red}{+4.5\%} & \cellcolor{gray!20}\textcolor{red}{+13.5\%} & \cellcolor{gray!20}\textcolor{red}{+30.2\%} & \cellcolor{gray!20}\textcolor{red}{+11.6\%} \\
    \bottomrule
  \end{tabular}
  }
  % \vspace{-5pt}
\end{table*}

\begin{table*}[t]
  \centering
  \caption{
  Quantitative comparison with broader satellite-to-radar retrieval
  baselines on SE China and \sevir{}.
  CSI@mod. denotes 20 dBZ for SE China and encoded VIL@160 for
  \sevir{}, while CSI@heavy denotes 35 dBZ and encoded VIL@219,
  respectively.
  % SE China MAE is reported on the dataset target scale, and Params denotes
  % trainable parameters.
  Time denotes SE China inference latency in ms/sample.
  The best score in each column is highlighted in
  \textcolor{green!60!black}{green}, and the second-best score is
  underlined.
  }
  \label{tab:external_baselines}

  \scriptsize
  \setlength{\tabcolsep}{2.15pt}
  \renewcommand{\arraystretch}{0.98}

  \resizebox{\textwidth}{!}{%
  \begin{tabular}{@{}l|c|*{7}{c}|*{5}{c}@{}}
    \toprule
    \multirow[c]{2}{*}{\textbf{Method}} &
    
    & \multicolumn{7}{c|}{SE China}
    & \multicolumn{5}{c}{\sevir{}} \\
    \cmidrule(lr){3-9}
    \cmidrule(lr){10-14}

    & Params (M)$\downarrow$
    & Time$\downarrow$
    & Avg.CSI$\uparrow$
    & CSI-M$\uparrow$
    & CSI-H$\uparrow$
    & Avg.HSS$\uparrow$
    & SSIM$\uparrow$
    & MAE$\downarrow$
    & Avg.CSI$\uparrow$
    & CSI-M$\uparrow$
    & CSI-H$\uparrow$
    & Avg.HSS$\uparrow$
    & LPIPS$\downarrow$ \\
    \midrule

    AA-TransUNet (2022)
    & 39.88
    & \second{6.7}
    & 0.189 & 0.234 & 0.075 & 0.275 & 0.539 & 2.980
    & 0.301 & \second{0.317} & 0.095 & 0.437 & 0.318 \\

    Earthformer (2022)
    & 8.68
    & 13.4
    & 0.206 & 0.264 & 0.060 & 0.297 & 0.555 & 2.864
    & 0.293 & 0.304 & 0.069 & 0.422 & 0.360
    \\

    SmaAt-UNet (2021)
    & 4.02
    & \best{3.6}
    & 0.161 & 0.203 & 0.070 & 0.233 & 0.501 & 3.498
    & 0.292 & 0.313 & 0.055 & 0.420 & 0.362
    \\

    DiffCast (2024)
    & 46.25
    & 3302.8
    & 0.223 & 0.263 & 0.106 & 0.322 & \best{0.590}
    & \best{2.681}
    & \best{0.310} & \best{0.321} & 0.126
    & \best{0.448} & 0.236 \\

    Pix2Pix (2024)
    & 31.36
    & 7.4
    & 0.229 & 0.267 & 0.132 & 0.334
    & 0.523 & 2.979
    & 0.240 & 0.269 & 0.014 & 0.355 & 0.349 \\

    MeanFlow (2025)
    & 107.80
    & 22.3
    & 0.197 & 0.241 & 0.092 & 0.292
    & 0.498 & 3.086
    & 0.200 & 0.177 & 0.044 & 0.299 & 0.300 \\

    Arrow (2025)
    & 3.36
    & 160.79
    & 0.198 & 0.232 & 0.097 & 0.295 & 0.549 & 3.101
    & 0.213 & 0.218 & 0.080 & 0.326 & 0.405 \\

    Weather-RF (2026)
    & \second{3.07}
    & 388.27
    & \second{0.265}
    & \second{0.315}
    & \second{0.142}
    & \second{0.385}
    & 0.550
    & 2.912
    & 0.282
    & 0.290
    & 0.123
    & 0.415
    & 0.206 \\

    LiteFM-UNet (2026)
    & 5.54
    & 416.30
    & 0.200 & 0.237 & 0.094 & 0.298 & 0.555 & 3.010
    & 0.299 & 0.311 & \second{0.133}
    & 0.435 & \second{0.201} \\

    \midrule

    \method{} (Ours)
    & \best{2.61}
    & 429.22
    & \best{0.270}
    & \best{0.318}
    & \best{0.146}
    & \best{0.392}
    & \second{0.578}
    & \second{2.725}
    & \second{0.306}
    & 0.315
    & \best{0.151}
    & \second{0.445}
    & \best{0.192} \\

    \bottomrule
  \end{tabular}%
  }
  % \vspace{-10pt}
\end{table*}

\subsection{Lightweight Encoder-Decoder}

To further reduce velocity-network cost while preserving encoder details, we replace the high-channel skip concatenation of conventional U-Nets with gated additive skip fusion.
This design complements the compact SLW blocks: SLW reduces feature-transformation cost, and the skip gate controls detail injection without increasing decoder channel width.
\method{} uses a compact encoder-decoder with channel multipliers $(1,2,4)$ and base width 40.
The input to the network is the channel-wise concatenation of the current noisy state and the satellite condition.
Each down block performs a $1\times1$ projection, one spectral-local-wavelet block, and depthwise downsampling.
The bottleneck contains two spectral-local-wavelet blocks, where spectral mixing is cheapest and has the largest effective field of view.
The decoder implements this skip fusion as
\begin{equation}
  \mathbf{F}_{\mathrm{up}} = P_x(U(\mathbf{F})),
  \quad
  \mathbf{F}_{\mathrm{skip}} = P_s(\mathbf{S}),
\end{equation}
\begin{equation}
  \mathbf{G} = \sigma(P_g([\mathbf{F}_{\mathrm{up}},\mathbf{F}_{\mathrm{skip}}])),
\end{equation}
\begin{equation}
  \mathbf{F}_{\mathrm{fused}} = \mathbf{F}_{\mathrm{up}} + \mathbf{G}\odot \mathbf{F}_{\mathrm{skip}}.
\end{equation}
where $\mathbf{S}$ is the encoder skip feature, $U$ is bilinear upsampling, $P_x$ and $P_s$ align channel width, and $P_g$ predicts the skip gate $\mathbf{G}$.
The skip gate controls encoder-detail injection during upsampling, while the branch gate selects operators inside each SLW block.
The main savings come from the narrow encoder-decoder, depthwise local operators, and additive gated skip fusion.

\subsection{Large-Area Tiled Inference Protocol}

For domains beyond the training crop, \method{} is evaluated convolutionally on overlapping tiles and assembled into a continuous retrieval field.
This is supported by its spatial operators, including convolution, Fourier convolution, Haar transform, and interpolation, without learned absolute positional embeddings or fixed-length token sequences.
In the China-scale case, tile predictions are merged by Hann-window weighted stitching; the tile size, overlap, and blending details are provided in the Appendix.
This protocol supports grid-size flexible inference within memory limits, but does not imply distribution-free generalization across regions, sensors, seasons, or precipitation regimes.

\begin{figure}[ht]
  \centering
  \includegraphics[width=\linewidth]{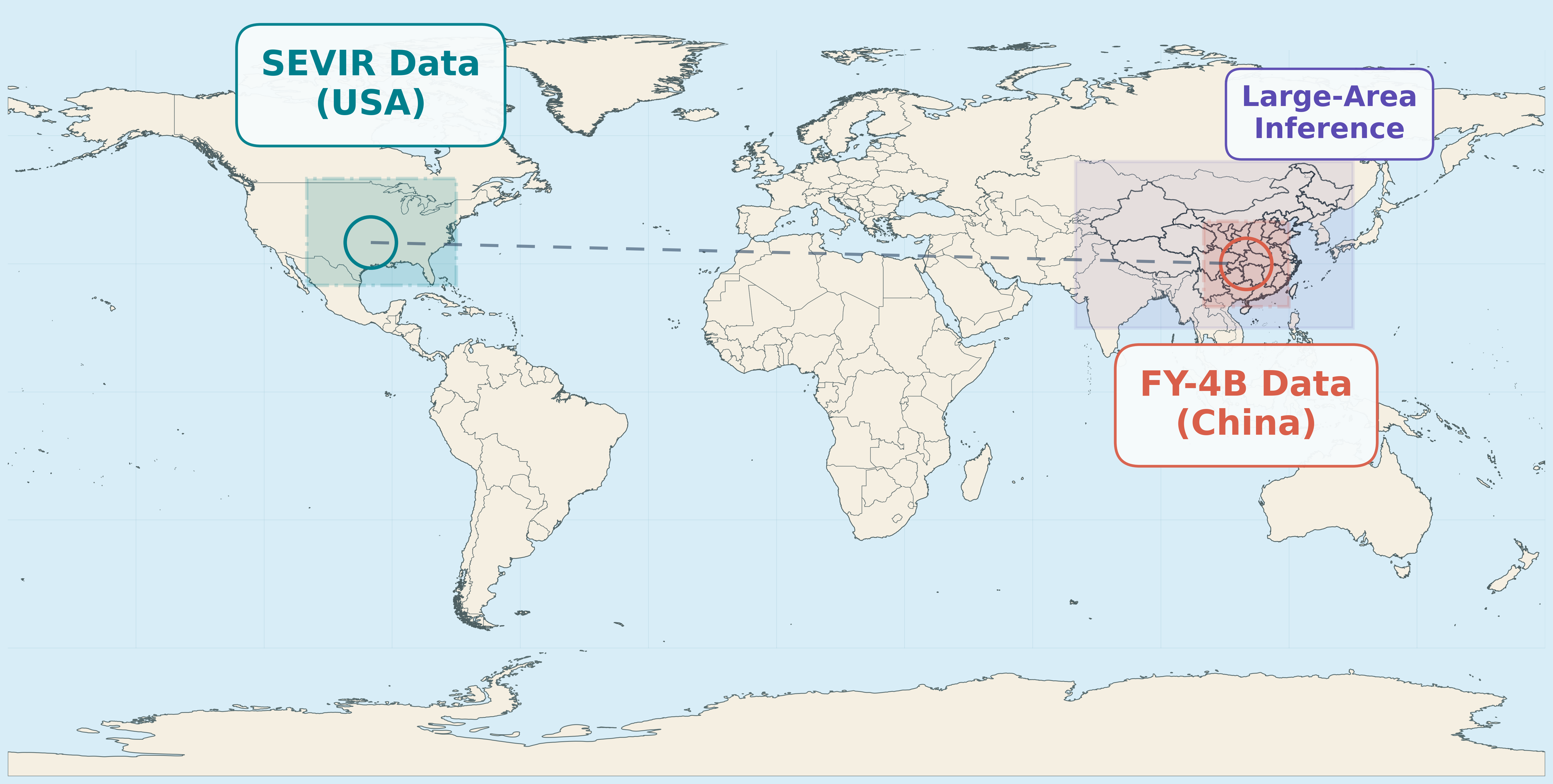}
  \caption{
  Study-region overview for the two satellite-to-radar retrieval settings.
  }
  \label{fig:study_regions}
  \vspace{-15pt}
\end{figure}
\begin{figure*}[t]
    \centering
    \includegraphics[width=\linewidth]{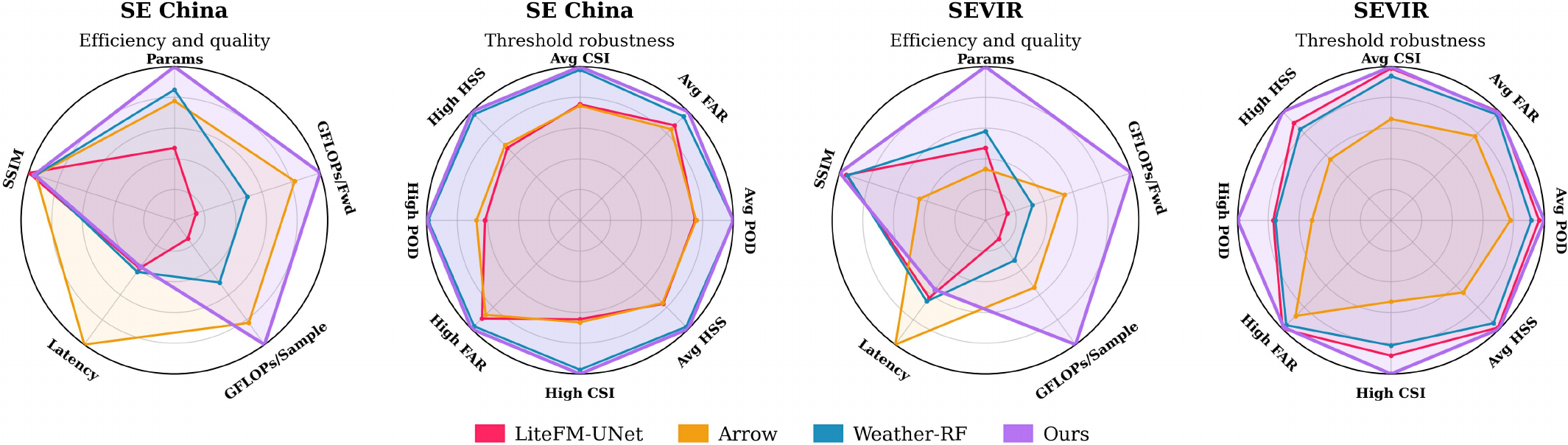}
    \caption{
    Comparison of efficiency, reconstruction quality, and threshold
    robustness on SE China and \sevir{}.
    The radar plots show the efficiency-quality trade-off and the average
    and high-intensity event detection performance for each dataset.
    All metrics are direction-aware normalized such that larger radial values consistently indicate better performance.
    }
    \label{fig:combined_radar}
    % \vspace{-10pt}
\end{figure*}
% \begin{figure*}[t]
%   \centering
%   \includegraphics[width=0.98\linewidth]{fig/f_vis_huadong_v1.pdf}
%   \caption{
%   Qualitative main-experiment comparison on the southeast China satellite-to-reflectivity dataset.
%   xxxx.
%   }
%   \label{fig:china_retrieval_case1}
%     \vspace{-2pt}
% \end{figure*}

\begin{figure*}[htbp]
  \centering
  \includegraphics[width=\linewidth]{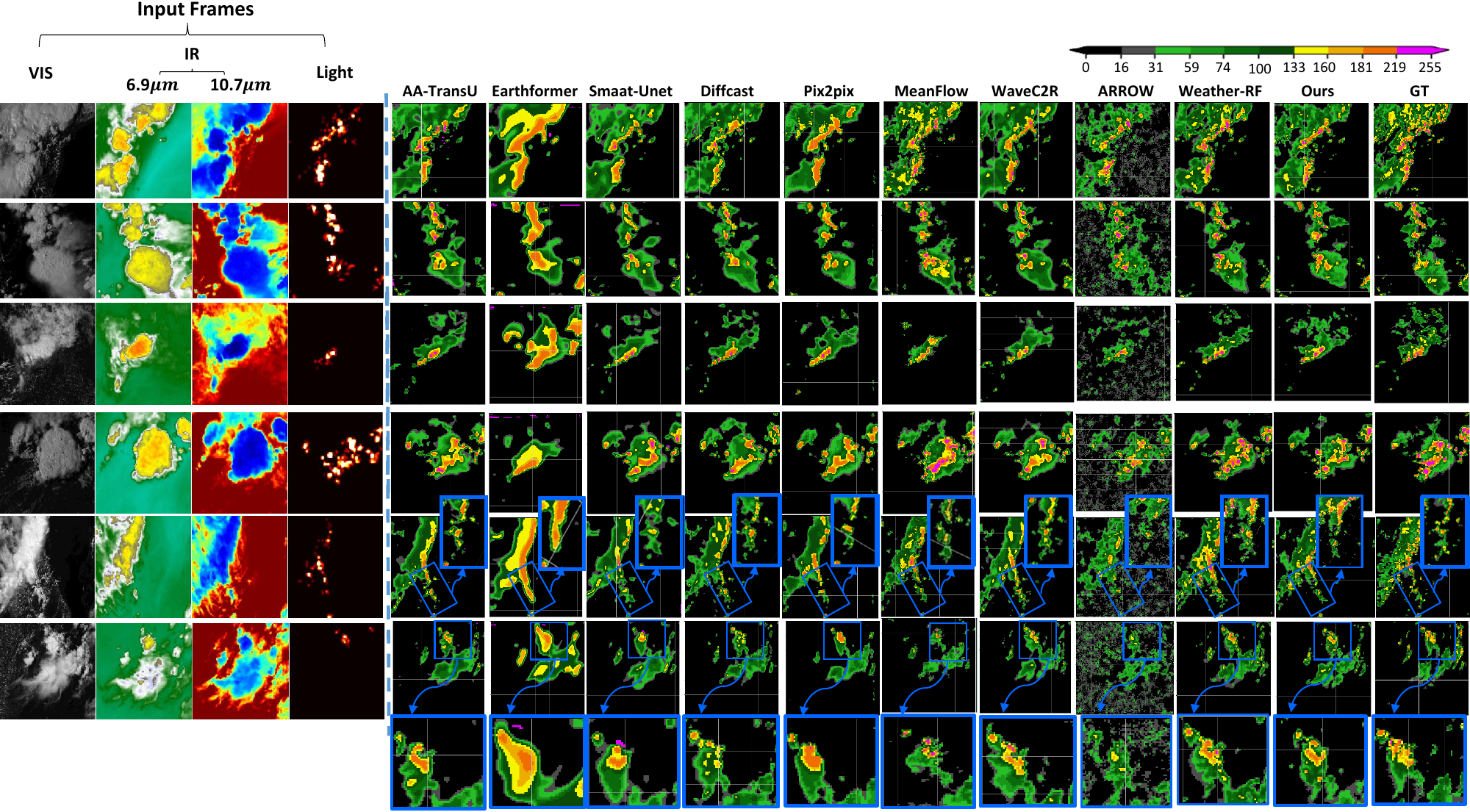}
  \caption{
  Qualitative main-experiment comparison on the \sevir{} satellite-to-VIL benchmark.
  The panels compare satellite conditions, ground-truth VIL, model predictions, and absolute-error maps for representative storm cases.
  They complement the threshold metrics by showing spatial displacement, intensity underestimation, and false precipitation spread.
  }
  \label{fig:sevir_retrieval_case1}
    % \vspace{-10pt}
\end{figure*}

\section{Experiments}
\label{sec:experiments}

\subsection{Datasets and Protocol}

\noindent\textbf{Datasets.}
We train and test \method{} on two satellite-based radar retrieval benchmarks: a southeast China satellite-to-reflectivity dataset and the public \sevir{} satellite-to-VIL dataset (Fig.~\ref{fig:study_regions}) \citep{veillette2020sevir}.
We further conduct large-area inference over China for deployment-oriented qualitative analysis.
Detailed dataset and inference settings are provided in the Appendix.

\noindent\textbf{Baselines.}
LiteFM-UNet serves as a controlled pixel-space flow baseline that shares the same objective and sampler with \method{} but uses a conventional U-Net velocity backbone.
We also compare with representative CNN/transformer architectures (AA-TransUNet~\citep{yang2022aa}, Earthformer~\citep{gao2022earthformer}, SmaAt-UNet~\citep{trebing2021smaat}), generative baselines (DiffCast~\citep{yu2024diffcast}, Pix2Pix~\citep{akter2024pix2pix}, MeanFlow~\citep{geng2025mean}), and recent weather-specific models (Arrow~\citep{tian2025arrow}, Weather-RF~\citep{schusterbauer2026probabilistic}).

\noindent\textbf{Evaluation Protocol.}
Following the evaluation protocol of WaveC2R~\citep{shi2026wavec2r}, we report image quality, threshold skill, qualitative cases, model size, hook-based FLOPs, and runtime; dataset details, thresholds, and implementation settings are provided in the Appendix.

\subsection{Main Comparison}

\noindent\textbf{Quantitative Comparison.}
Table~\ref{tab:main_metrics} compares \method{} with LiteFM-UNet to isolate the effect of replacing the U-Net velocity backbone under the same pixel-space flow objective and sampler.
Across both datasets, \method{} markedly reduces parameters and sampling FLOPs while maintaining or improving reconstruction quality, especially at heavy-precipitation thresholds where localized convective cores are most sensitive to smoothing.
Table~\ref{tab:external_baselines} further shows a favorable quality-efficiency balance against representative CNN, transformer, GAN, diffusion, flow, and weather-specific baselines, while using the fewest parameters among the generative baselines.
These results indicate that lightweight velocity parameterization can preserve pixel-space retrieval fidelity while improving the quality-efficiency trade-off for radar-like precipitation reconstruction.

\begin{figure*}[ht]
  \centering
  \includegraphics[width=\linewidth]{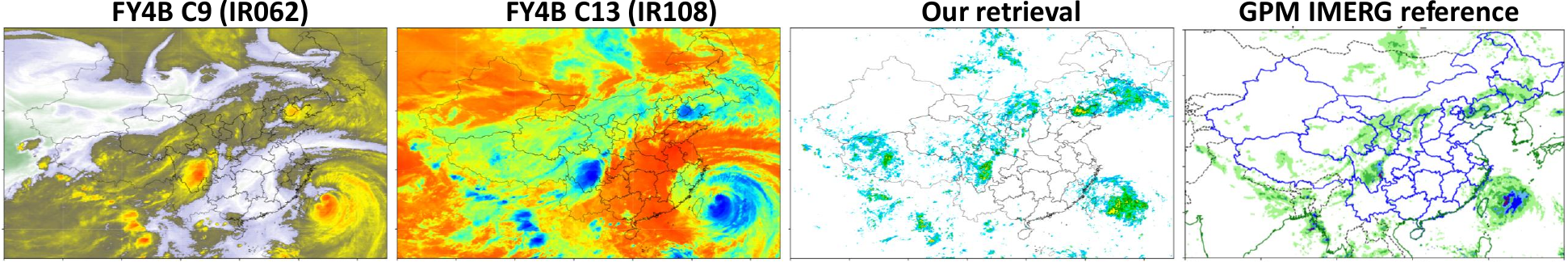}
  % \caption{
  % Large-area Typhoon Bavi case over China at 00:00 UTC on 11 July 2026.
  % From left to right: FY-4B IR062, FY-4B IR108, WaveOp-LiteFM retrieval,
  % and GPM IMERG reference.
  % }
  \caption{
  Large-area Typhoon Bavi case over China at 00:00 UTC on 11 July 2026, spanning $70.0^\circ$E--$135.0^\circ$E and $15.0^\circ$N--$54.0^\circ$N.
  From left to right: FY-4B IR062, FY-4B IR108, WaveOp-LiteFM retrieval,
  and GPM IMERG reference.
  }
  \label{fig:large_area_gpm}
    % \vspace{-5pt}
\end{figure*}
\begin{figure}[htb]
  \centering
  \includegraphics[width=\linewidth]{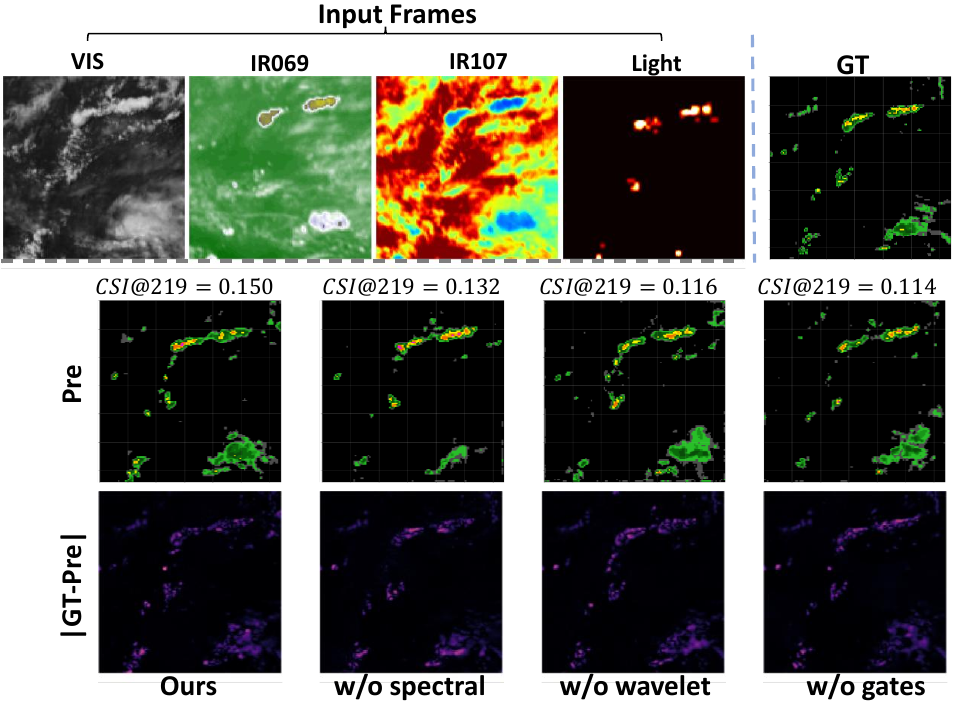} 
  \caption{
  Qualitative ablation study on the \sevir{} dataset.
  Retrieval and absolute-error maps are shown for the same storm event under the full model and its ablated variants. 
  }
  \label{fig:ablation_case}
  % \vspace{-10pt}
\end{figure}

% \begin{table*}[t]
%   \centering
%   \caption{
%   \sevir{} ablation results for \method{}.
%   % Weather-RF and Model 1 LiteFM-UNet are included as reference baselines; the remaining rows evaluate the spectral branch, wavelet shrinkage branch, and adaptive gated fusion.
%   % Avg.CSI and Avg.HSS average over encoded VIL thresholds 74, 133, 160, 181, and 219.
%   }
%   \label{tab:ablation_design}
%   \scriptsize
%   \setlength{\tabcolsep}{2.4pt}
%   \renewcommand{\arraystretch}{1.08}
%   \resizebox{\textwidth}{!}{
%   \begin{tabular}{lcccccccccc}
%     \toprule
%     Variant & Params & GFLOPs/sample & Time (ms) $\downarrow$ & PSNR $\uparrow$ & SSIM $\uparrow$ & LPIPS $\downarrow$ & Avg.CSI $\uparrow$ & Avg.HSS $\uparrow$ & CSI@219 $\uparrow$ & HSS@219 $\uparrow$ \\
%     \midrule
%     % Weather-RF & 4.51M & 307.03 & 30.83 & 20.90 & 0.518 & 0.206 & 0.282 & 0.415 & 0.123 & 0.218 \\
%     % LiteFM-UNet (M1) & 5.54M & 664.79 & 32.25 
%     % & 21.25 & 0.523 & 0.201 & 0.299 & 0.435 & 0.133 & 0.233 \\
%     % \midrule
%     \method{} & 2.61M & 99.15 & 36.03 
%     & 21.32 & 0.548 & 0.192 & 0.306 & 0.445 & 0.151 & 0.260 \\
%     w/o spectral & 1.11M & 94.43 & 35.48 
%     & 20.53 & 0.528 & 0.178 & 0.302 & 0.440 & 0.132 & 0.238 \\
%     w/o wavelet & 2.37M & 71.89 & 27.90 
%     & 21.22 & 0.529 & 0.197 & 0.285 & 0.418 & 0.116 & 0.207 \\
%     w/o gates & 2.54M & 86.57 & 39.67 
%     & 21.43 & 0.518 & 0.196 & 0.293 & 0.427 & 0.114 & 0.203 \\
%     \bottomrule
%   \end{tabular}
%   }
%   \vspace{-10pt}
% \end{table*}
\begin{table*}[t]
  \centering
  \caption{
  % \sevir{} ablation results for \method{}.
  % The best score in each column is highlighted in \textcolor{green!60!black}{green}.
  % % Weather-RF and Model 1 LiteFM-UNet are included as reference baselines; the remaining rows evaluate the spectral branch, wavelet shrinkage branch, and adaptive gated fusion.
  % % Avg.CSI and Avg.HSS average over encoded VIL thresholds 74, 133, 160, 181, and 219.
  Ablation study on the \sevir{} dataset.
  The best score is highlighted in \textcolor{green!60!black}{green}, and the second-best score is underlined.
  }
  \label{tab:ablation_design}
  \scriptsize
  \setlength{\tabcolsep}{2.4pt}
  \renewcommand{\arraystretch}{1.08}
  \resizebox{\textwidth}{!}{
  \begin{tabular}{lccc|ccccccc}
    \toprule
    Variant & Params & GFLOPs/sample & Time (ms) $\downarrow$ & PSNR $\uparrow$ & SSIM $\uparrow$ & LPIPS $\downarrow$ & Avg.CSI $\uparrow$ & Avg.HSS $\uparrow$ & CSI@219 $\uparrow$ & HSS@219 $\uparrow$ \\
    \midrule
    % Weather-RF & 4.51M & 307.03 & 30.83 & 20.90 & 0.518 & 0.206 & 0.282 & 0.415 & 0.123 & 0.218 \\
    % LiteFM-UNet (M1) & 5.54M & 664.79 & 32.25 
    % & 21.25 & 0.523 & 0.201 & 0.299 & 0.435 & 0.133 & 0.233 \\
    % \midrule
    \method{} (Ours) & 2.61M & 99.15 & 36.03 
    & \second{21.32} & \best{0.548} & \second{0.192} & \best{0.306} & \best{0.445} & \best{0.151} & \best{0.260} \\
    w/o spectral & 1.11M & 94.43 & 35.48 
    & 20.53 & 0.528 & \best{0.178} & \second{0.302} & \second{0.440} & \second{0.132} & \second{0.238} \\
    w/o wavelet & 2.37M & 71.89 & 27.90 
    & 21.22 & \second{0.529} & 0.197 & 0.285 & 0.418 & 0.116 & 0.207 \\
    w/o gates & 2.54M & 86.57 & 39.67 
    & \best{21.43} & 0.518 & 0.196 & 0.293 & 0.427 & 0.114 & 0.203 \\
    \bottomrule
  \end{tabular}
  }
  \vspace{-10pt}
\end{table*}

\noindent\textbf{Threshold Skill and Qualitative Retrieval.}
Because radar retrieval is dominated by weak or non-precipitating pixels, average image metrics can obscure performance on convective cores.
Fig.~\ref{fig:combined_radar} therefore summarizes threshold skill and efficiency, with the full CSI, FAR, POD, and HSS values reported in the Appendix.
\method{} maintains strong high-threshold skill while preserving a compact efficiency profile, indicating that the lightweight backbone does not simply trade heavy-precipitation fidelity for lower cost.

Fig.~\ref{fig:sevir_retrieval_case1} shows representative \sevir{} cases.
Compared with the baselines, \method{} better preserves compact precipitation cores and sharp echo boundaries, consistent with the high-threshold skill reported above.

\subsection{Efficiency}

For flow-based retrieval, inference cost is dominated by repeated velocity-network evaluations during Euler sampling.
In the controlled comparison, \method{} uses 52.9\% fewer trainable parameters than LiteFM-UNet and reduces hook-based sampling FLOPs by about $6.7\times$ on both SE China and \sevir{} (Table~\ref{tab:main_metrics}).
These gains reflect the compact SLW backbone and gated additive skip fusion; Fig.~\ref{fig:combined_radar} summarizes the quality-efficiency trade-off.
Additional quantitative efficiency metrics are provided in the Appendix.

\subsection{China-scale Large-Area Inference Case}

Finally, we assess whether the trained model can be applied to a domain larger than the training crop without apparent tiling artifacts.
Fig.~\ref{fig:large_area_gpm} shows a China-scale Typhoon Bavi case generated by overlapping tiled inference and Hann-window weighted stitching.
The FY-4B infrared inputs, stitched radar-like retrieval, and GPM IMERG reference exhibit broadly consistent large-scale precipitation organization.
Because GPM IMERG and the retrieval target differ in sensing physics and product definition, this case is used as a qualitative deployment check rather than a pixel-wise benchmark.

\subsection{Ablation Studies}

To verify the role of each architectural component, we conduct controlled ablations on the public \sevir{} benchmark.
Table~\ref{tab:ablation_design} and Fig.~\ref{fig:ablation_case} show that the complete \method{} achieves the strongest threshold skill and the best SSIM among the ablated variants, while retaining a compact sampling profile.
The wavelet branch is most important for high-frequency radar details and intense-core skill, while adaptive gates improve robustness across precipitation morphologies.
Although removing the spectral branch lowers parameter count and LPIPS, it weakens high-threshold skill, supporting the role of spectral mixing in storm-scale organization.
Detailed ablation settings are provided in the Appendix.

\section{Conclusion}
In this work, we address the compute bottleneck of pixel-space generative satellite-to-radar retrieval with \method{}, a lightweight neural-operator flow matching framework.
\method{} preserves the stable pixel-space conditional flow objective and replaces the heavy U-Net velocity backbone with spectral-local-wavelet operator blocks.
The spectral branch models storm-scale organization, the depthwise local branch captures short-range convective interactions, and the wavelet branch emphasizes high-frequency echo boundaries and radar details.
Adaptive branch gating and gated additive skip fusion further reduce redundant processing while preserving encoder information.
Experiments on the public \sevir{} and southeast China benchmarks show that \method{} achieves strong radar retrieval skill and perceptual quality with lower sampling cost.

\clearpage
\bibliographystyle{IEEEtran}
\bibliography{egbib}

\clearpage
\appendix
\label{appendix}

This supplementary material provides additional dataset and baseline details in Sec.~\ref{app:dataset_baselines}, implementation and training details in Sec.~\ref{app:implementation}, and evaluation details in Sec.~\ref{app:evaluation_protocol}.
It also includes qualitative retrieval and large-area deployment cases in Secs.~\ref{app:china_cases} and~\ref{app:large_area_case}, together with diagnostic figures, detailed threshold and efficiency results, ablation analysis, and tiled-inference details in Secs.~\ref{app:visual_diagnostics}, \ref{app:detailed_results}, \ref{app:ablation_design}, and~\ref{app:tiled_inference}.

\subsection{Datasets, Baselines, and Study Regions}
\label{app:dataset_baselines}

\noindent\textbf{Baselines.}
The controlled baseline is LiteFM-UNet, which uses the same pixel-space conditional flow objective, 20-step Euler sampler, data splits, and evaluation code as \method{}, but retains a conventional U-Net velocity backbone.
This comparison isolates the effect of replacing the velocity network with the proposed lightweight spectral-local-wavelet operator backbone.
For broader context, we also compare with representative CNN and transformer retrieval models, generative baselines, one-step flow models, and recent weather-specific models, including Arrow and Weather-RF.

\noindent\textbf{Datasets.}
We evaluate \method{} on two satellite-to-radar retrieval benchmarks.
The southeast China benchmark uses FY-4B infrared satellite observations as conditions and radar-derived reflectivity-like fields as targets on the native $500\times500$ regional grid.
The \sevir{} benchmark uses visible, infrared, and lightning observations to retrieve VIL at $128\times128$ resolution, following the public storm-event setting.
Both benchmarks use the same conditional flow formulation and deterministic Euler sampler unless otherwise specified.
Table~\ref{tab:data_config} summarizes these dataset settings.

\begin{figure*}[ht]
  \centering
  \includegraphics[width=0.9\linewidth]{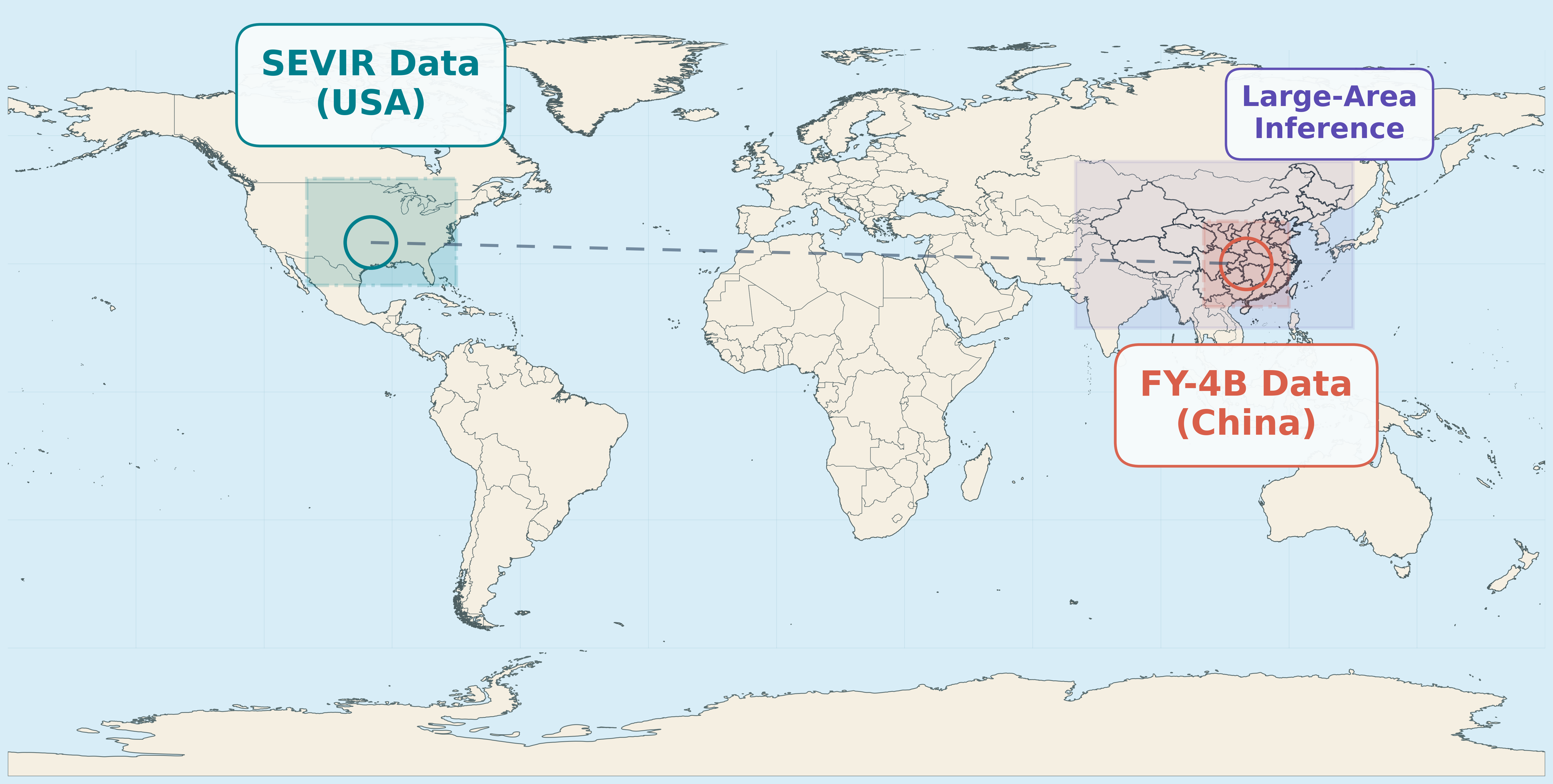}
  \caption{
  Appendix study-region overview for the public \sevir{} benchmark, the Southeast China regional benchmark ($100.0^\circ$E--$120.0^\circ$E, $20.0^\circ$N--$40.0^\circ$N), and the China-scale inference domain ($70.0^\circ$E--$135.0^\circ$E, $15.0^\circ$N--$54.0^\circ$N).
  The Southeast China setting uses FY-4B infrared satellite channels to retrieve a reflectivity-like precipitation field on a native regional grid, while the \sevir{} setting uses visible, infrared, and lightning channels over the United States to retrieve VIL.
  The China-scale domain is used for deployment-oriented tiled inference.
  }
  \label{fig:supp_study_regions}
\end{figure*}

\noindent\textbf{Study regions.}
Figure~\ref{fig:supp_study_regions} summarizes the three spatial domains used in the paper.
The \sevir{} domain provides a public U.S. storm-event benchmark for multi-source satellite-to-VIL retrieval.
The southeast China domain ($100.0^\circ$E--$120.0^\circ$E, $20.0^\circ$N--$40.0^\circ$N) evaluates FY-4B-to-radar retrieval in the regional setting used for the main benchmark, where preserving localized convective echoes on the native grid is the primary concern.
The China-scale domain is used only for deployment-oriented qualitative inference: the FY-4B mosaic spans $70.0^\circ$E--$135.0^\circ$E and $15.0^\circ$N--$54.0^\circ$N at $0.04^\circ$ spacing, which is substantially larger than the training crop and is processed by overlapping tiled inference.

\subsection{Implementation and Training Protocol}
\label{app:implementation}
\label{app:training_protocol}

\noindent\textbf{\method{} configuration.}
All \method{} experiments use the same velocity backbone.
The backbone has base width 40, channel multipliers $(1,2,4)$, a 192-dimensional flow-time embedding, Fourier mode budget $(10,10)$, spectral ratio 4, gated branch fusion, and gated additive skip fusion.
The spectral branch is enabled from the second encoder level onward, so the most expensive high-resolution feature maps are handled by local and wavelet operators while global mixing is applied at lower spatial resolutions.
The Haar wavelet branch is enabled in the full model and removed only in the corresponding ablation.

\begin{table*}[htbp]
  \centering
  \caption{
  Dataset and configuration summary for the Model 4 experiments.
  The two datasets share the same \method{} training code and differ only in condition channels, target product, and evaluation scale.
  }
  \label{tab:data_config}
  \small
  \setlength{\tabcolsep}{4pt}
  \renewcommand{\arraystretch}{1.08}
  \begin{tabular}{lccccc}
    \toprule
    Dataset & Conditions & Target & Eval. scale & Image size & Split protocol \\
    \midrule
    SE China & IR069, IR107 & reflectivity-like field & 0--70 dBZ & native $500\times500$ & JSON split files \\
    \sevir{} & VIS, IR069, IR107, LGHT & VIL & 0--255 encoded VIL & $128\times128$ & date-based split \\
    \bottomrule
  \end{tabular}
\end{table*}

\begin{figure*}[ht]
  \centering
  \includegraphics[width=0.9\linewidth]{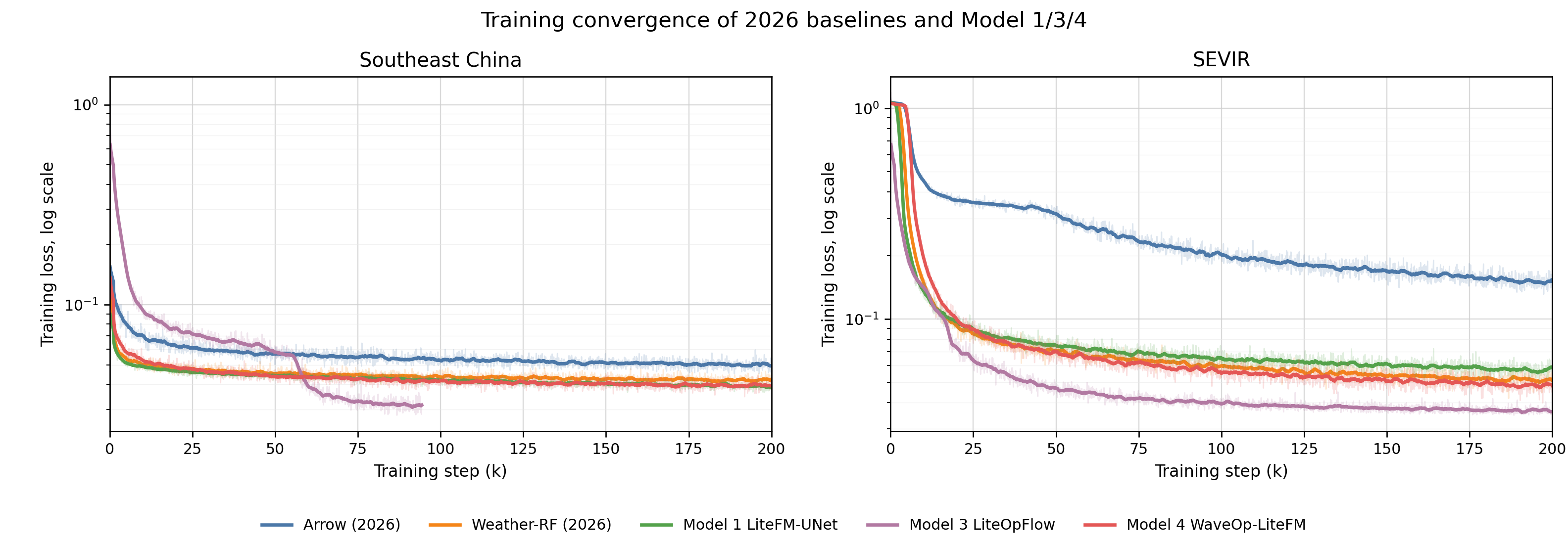}
  \caption{
  Training-loss convergence of the two 2026 baselines, Arrow and Weather-RF, together with LiteFM-UNet, LiteOpFlow, and the proposed \method{}.
  LiteOpFlow denotes the Model 3 latent/operator-flow variant considered during model development.
  Curves are shown for the southeast China and \sevir{} settings.
  The thick curves are smoothed training losses and the faint curves show the logged raw losses, using a logarithmic vertical axis.
  The plot is intended as an optimization diagnostic rather than a validation-performance curve.
  }
  \label{fig:supp_training_curves}
\end{figure*}

\begin{table*}[ht]
  \centering
  \footnotesize
  \setlength{\tabcolsep}{3pt}
  \renewcommand{\arraystretch}{1.12}
  \caption{Appendix testing thresholds and reported metric groups.}
  \label{tab:supp_eval_thresholds}
  \begin{tabular}{p{0.14\textwidth}p{0.18\textwidth}p{0.24\textwidth}p{0.22\textwidth}p{0.16\textwidth}}
    \toprule
    Dataset & Evaluation scale & Thresholds & Image metrics & Threshold metrics \\
    \midrule
    Southeast China & 0--70 dBZ target scale at native $500\times500$ resolution & 10, 20, 25, 30, and 35 dBZ & MSE, MAE, RMSE, PSNR, SSIM, LPIPS & CSI, POD, FAR, HSS at each threshold \\
    \sevir{} & Encoded VIL target scale at $128\times128$ resolution & 16, 74, 133, 160, 181, and 219 encoded VIL units & MSE, MAE, RMSE, PSNR, SSIM, LPIPS & CSI, POD, FAR, HSS at each threshold \\
    \bottomrule
  \end{tabular}
\end{table*}

\noindent\textbf{Training protocol.}
For the reported \method{} runs on both datasets, the model is trained with AdamW using learning rate $2\times10^{-4}$, weight decay $10^{-4}$, betas $(0.9,0.95)$, and gradient clipping at 1.0.
Training runs for 200k optimization steps in the reported final experiments.
Samples are generated periodically during training for qualitative monitoring, and checkpoints are saved every 5k steps.

The conditional flow path uses a linear interpolation between Gaussian noise and the radar target.
The flow-time variable is clipped with $\epsilon=10^{-4}$ for numerical stability, and no additional interpolation noise is added in the current configuration.
At inference, all reported deterministic evaluations use 20 Euler steps unless stated otherwise.
The optimization behavior of LiteFM-UNet, LiteOpFlow, \method{}, and the two 2026 baselines is shown in Figure~\ref{fig:supp_training_curves}, which is used only as a convergence diagnostic.

\subsection{Evaluation Protocol}
\label{app:evaluation_protocol}

Evaluation follows the image-quality and threshold-skill protocol used in WaveC2R~\citep{shi2026wavec2r} and is performed on fixed held-out splits for each dataset.
The evaluation code reports image-level metrics, threshold contingency scores, model size, FLOPs estimates, and wall-clock inference time.
For southeast China, predictions are mapped to a 0--70 dBZ scale and evaluated at thresholds $\{10,20,25,30,35\}$.
For \sevir{}, predictions are mapped to the encoded VIL scale and evaluated at thresholds $\{16,74,133,160,181,219\}$.
Table~\ref{tab:supp_eval_thresholds} lists the dataset-specific scaling conventions, thresholds, and metric groups.
Let $\hat{Y}$ and $Y$ denote the scaled prediction and target fields on a common evaluation grid, and let $\Omega$ be the set of valid pixels with $N=|\Omega|$.
For the southeast China setting, $\hat{Y}=70\hat{y}$ and $Y=70y$ are interpreted on the dBZ-like scale.
For \sevir{}, $\hat{Y}=255\hat{y}$ and $Y=255y$ are interpreted on the encoded VIL scale.
The image-level error metrics are computed over $\Omega$ as
\begin{align}
  \mathrm{MSE}
  &=
  \frac{1}{N}\sum_{p\in\Omega}\left(\hat{Y}_p-Y_p\right)^2, \\
  \mathrm{MAE}
  &=
  \frac{1}{N}\sum_{p\in\Omega}\left|\hat{Y}_p-Y_p\right|, \\
  \mathrm{RMSE}
  &=
  \sqrt{\mathrm{MSE}} .
\end{align}
Peak signal-to-noise ratio is reported as
\begin{equation}
  \mathrm{PSNR}
  =
  10\log_{10}
  \left(
  \frac{R^2}{\mathrm{MSE}}
  \right),
\end{equation}
where $R$ is the dynamic range of the evaluation scale, namely $R=70$ for southeast China and $R=255$ for \sevir{}.
SSIM and LPIPS are computed on the normalized $[0,1]$ fields used by the neural model, so that structural similarity and perceptual distance are not affected by the dataset-specific reporting scale.

For a threshold $\tau$, binary event fields are obtained from the scaled variables:
\begin{equation}
  \hat{E}_p^{(\tau)}=\mathbf{1}\{\hat{Y}_p\ge\tau\},\qquad
  E_p^{(\tau)}=\mathbf{1}\{Y_p\ge\tau\}.
\end{equation}
The contingency counts are
\begin{align}
  H_\tau &= \sum_{p\in\Omega}\mathbf{1}\{\hat{E}_p^{(\tau)}=1,E_p^{(\tau)}=1\}, \\
  M_\tau &= \sum_{p\in\Omega}\mathbf{1}\{\hat{E}_p^{(\tau)}=0,E_p^{(\tau)}=1\}, \\
  F_\tau &= \sum_{p\in\Omega}\mathbf{1}\{\hat{E}_p^{(\tau)}=1,E_p^{(\tau)}=0\}, \\
  C_\tau &= \sum_{p\in\Omega}\mathbf{1}\{\hat{E}_p^{(\tau)}=0,E_p^{(\tau)}=0\}.
\end{align}
Here $H_\tau$, $M_\tau$, $F_\tau$, and $C_\tau$ denote hits, misses, false alarms, and correct negatives.
The threshold scores are then
\begin{align}
  \mathrm{CSI}_\tau
  &=
  \frac{H_\tau}{H_\tau+M_\tau+F_\tau}, \\
  \mathrm{POD}_\tau
  &=
  \frac{H_\tau}{H_\tau+M_\tau}, \\
  \mathrm{FAR}_\tau
  &=
  \frac{F_\tau}{H_\tau+F_\tau}, \\
  \mathrm{HSS}_\tau
  &=
  \frac{2(H_\tau C_\tau-M_\tau F_\tau)}
  {(H_\tau+M_\tau)(M_\tau+C_\tau)+(H_\tau+F_\tau)(F_\tau+C_\tau)} .
\end{align}
Lower values are better for MSE, MAE, RMSE, LPIPS, FAR, parameter count, FLOPs, and inference time; higher values are better for PSNR, SSIM, CSI, POD, HSS, and throughput.

\begin{figure*}[htbp]
  \centering
  \includegraphics[width=0.9\linewidth]{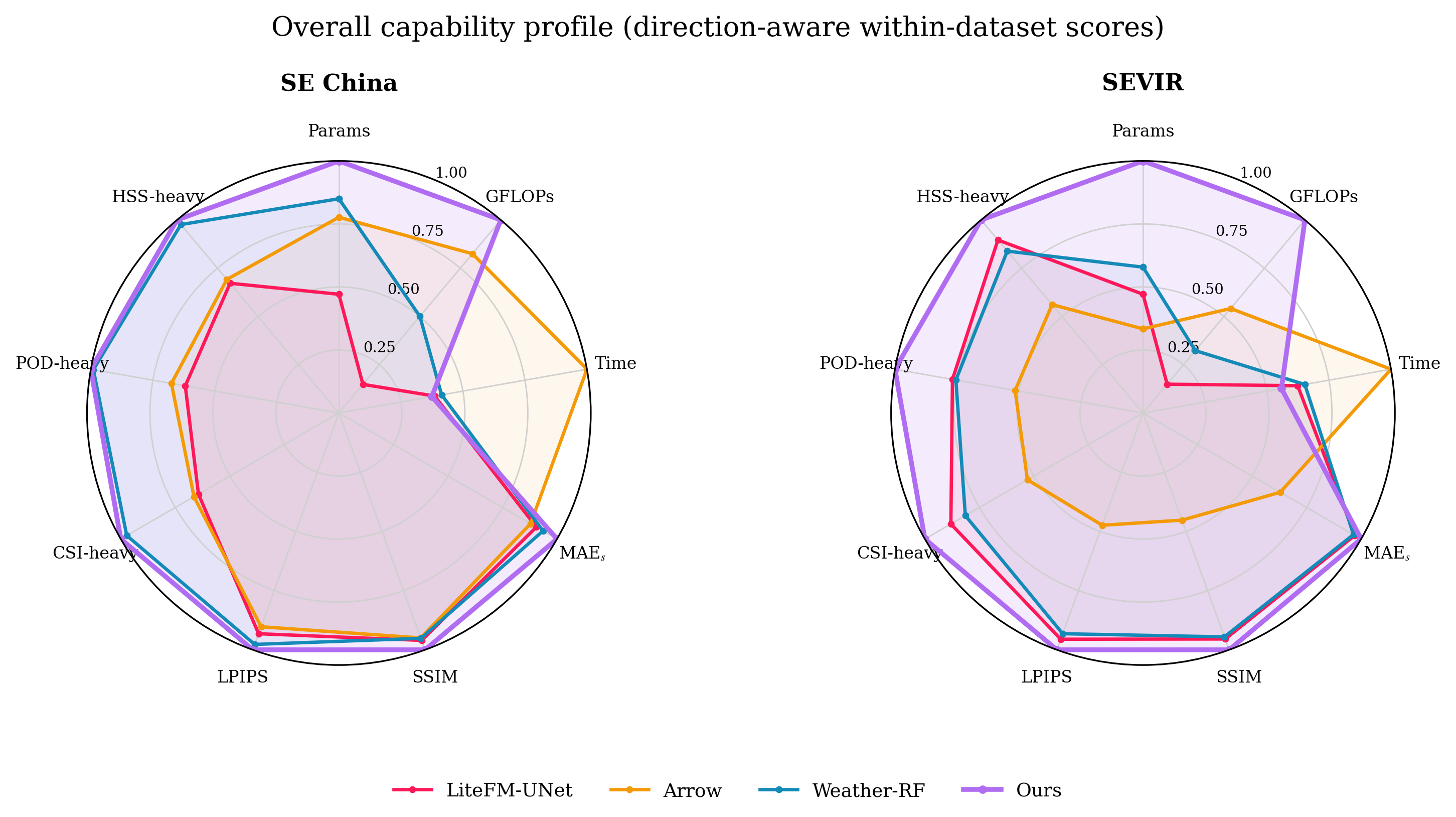}
  \caption{
  Appendix multi-metric capability profiles for Arrow, Weather-RF, Model 1 LiteFM-UNet, and Model 4 \method{}.
  Each axis is normalized within the corresponding dataset so that larger values indicate better performance or lower cost.
  }
  \label{fig:supp_radar_capability}
\end{figure*}

\begin{figure*}[htbp]
  \centering
  \includegraphics[width=0.9\linewidth]{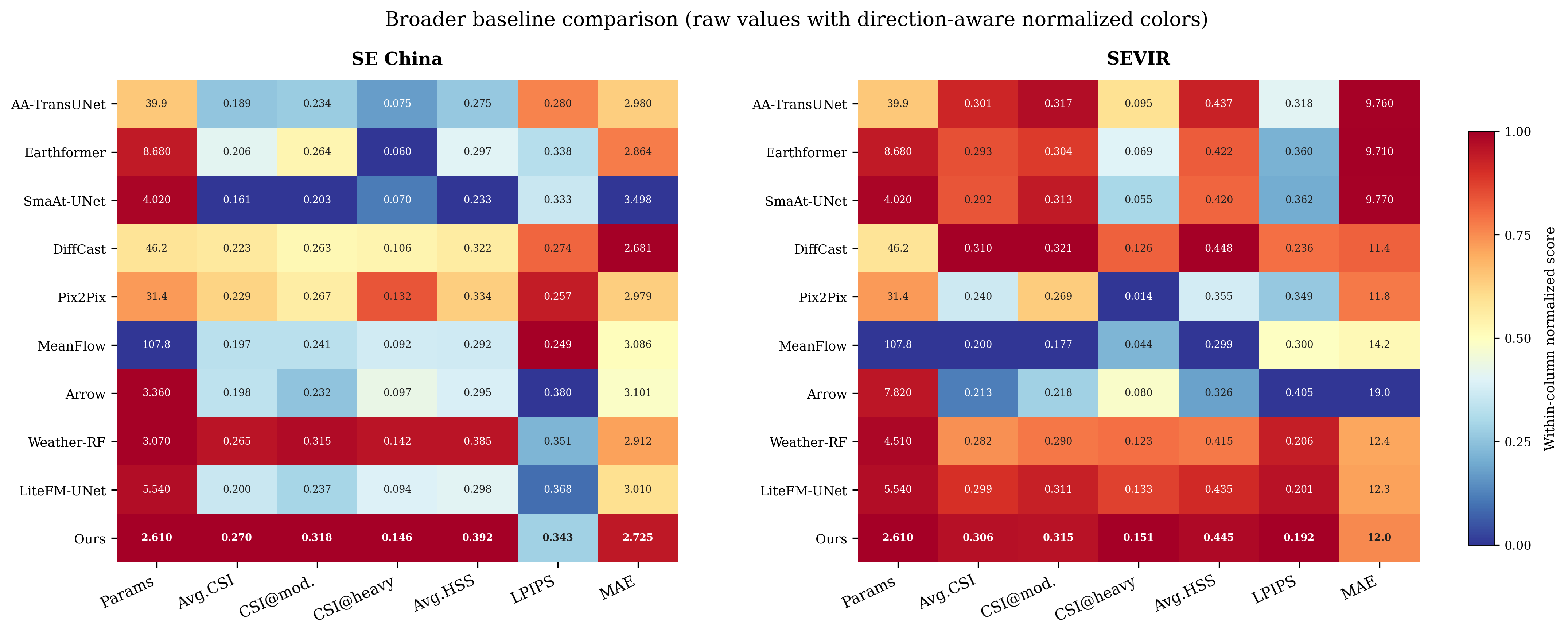}
  \caption{
  Appendix heatmap for the broader baseline comparison on the southeast China and \sevir{} settings.
  Cell text gives the raw metric value, while colors are normalized within each metric column with the correct direction of preference.
  This visualization complements the numerical comparison by making the relative strengths of the external baselines, LiteFM-UNet, and \method{} easier to inspect across accuracy, threshold skill, perceptual quality, and parameter count.
  }
  \label{fig:supp_external_baseline_heatmap}
\end{figure*}

\begin{figure*}[htbp]
  \centering
  \includegraphics[width=0.9\linewidth]{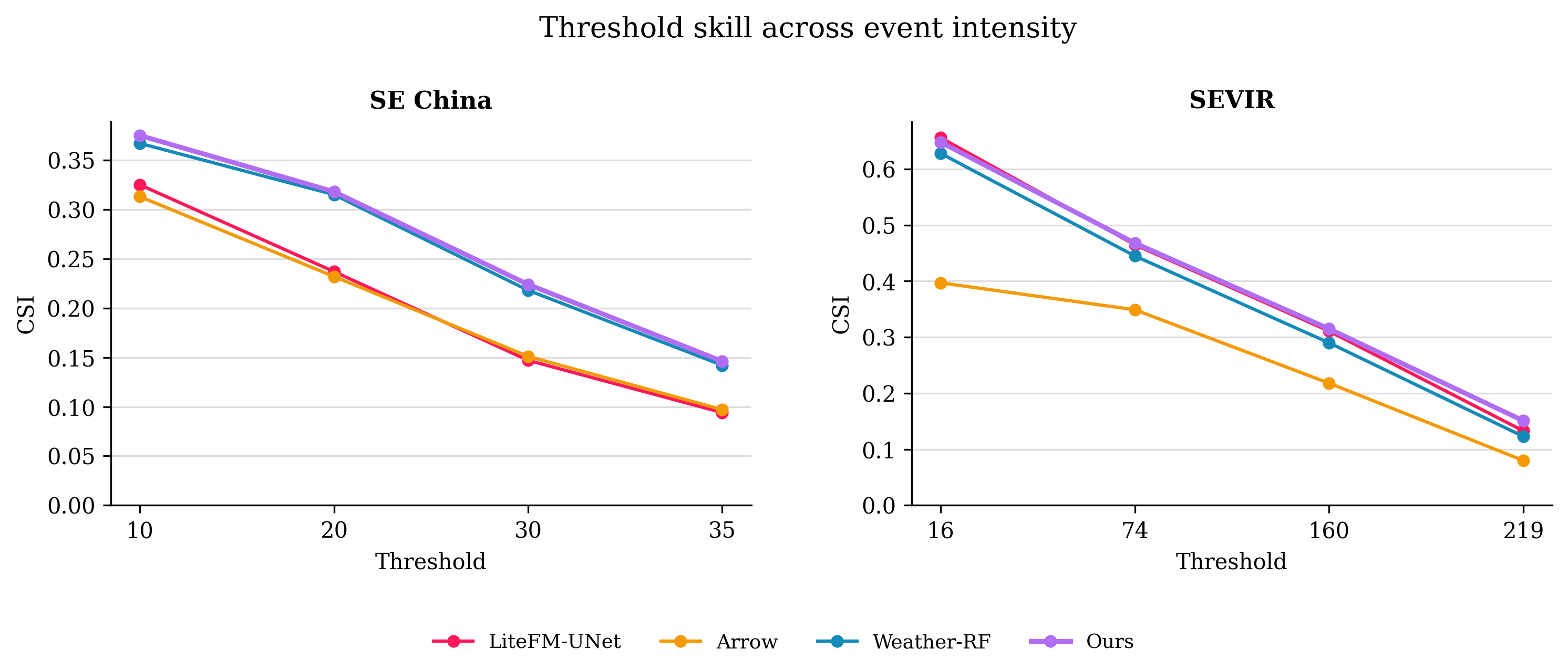}
  \caption{
  Appendix threshold-skill curves.
  The curves show raw CSI values at the evaluated precipitation thresholds for the southeast China and \sevir{} settings.
  This view complements the threshold-score table by showing how skill changes from light to intense precipitation levels.
  }
  \label{fig:supp_threshold_curves}
\end{figure*}

\begin{figure*}[htbp]
  \centering
  \includegraphics[width=0.9\linewidth]{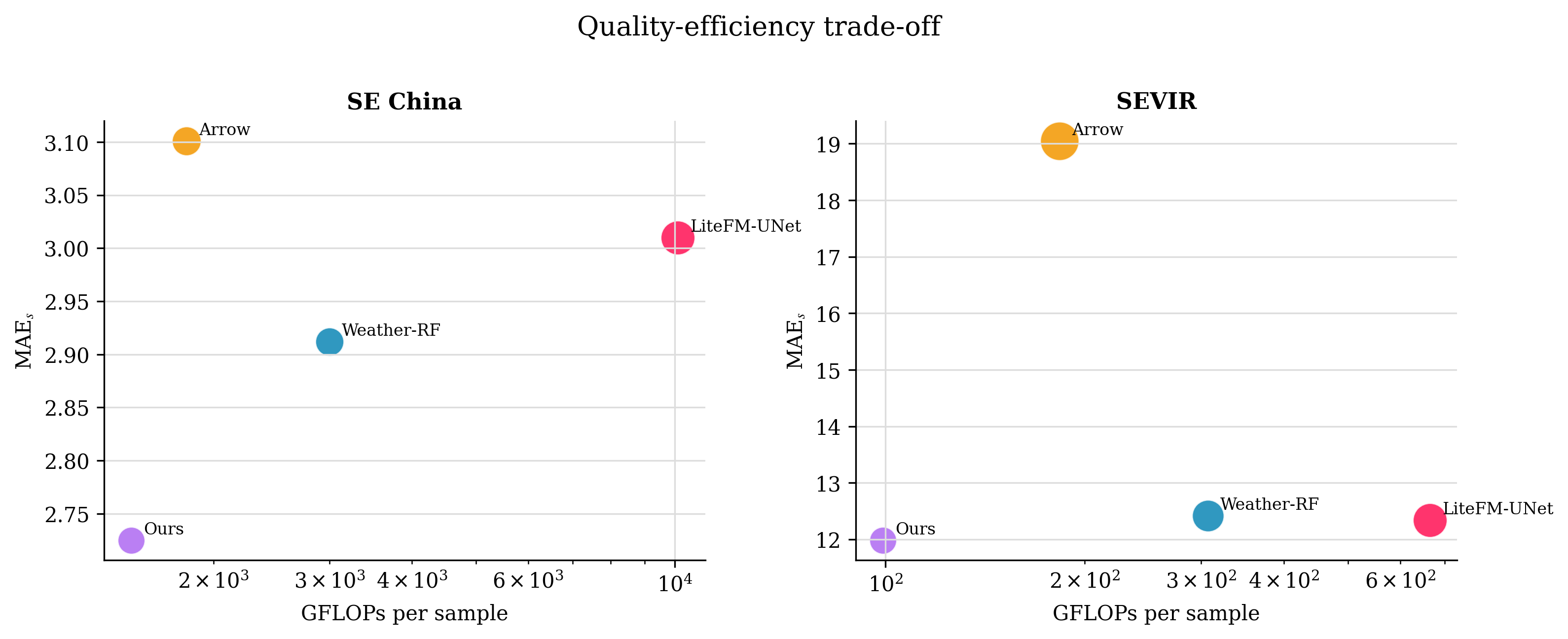}
  \caption{
  Appendix quality-efficiency trade-off.
  Each point compares RMSE against per-sample computational cost, with marker size proportional to parameter count.
  This view makes the efficiency motivation behind Model 4 visible alongside the 2026 baselines.
  }
  \label{fig:supp_quality_efficiency}
\end{figure*}

\begin{figure*}[htbp]
  \centering
  \includegraphics[width=\linewidth]{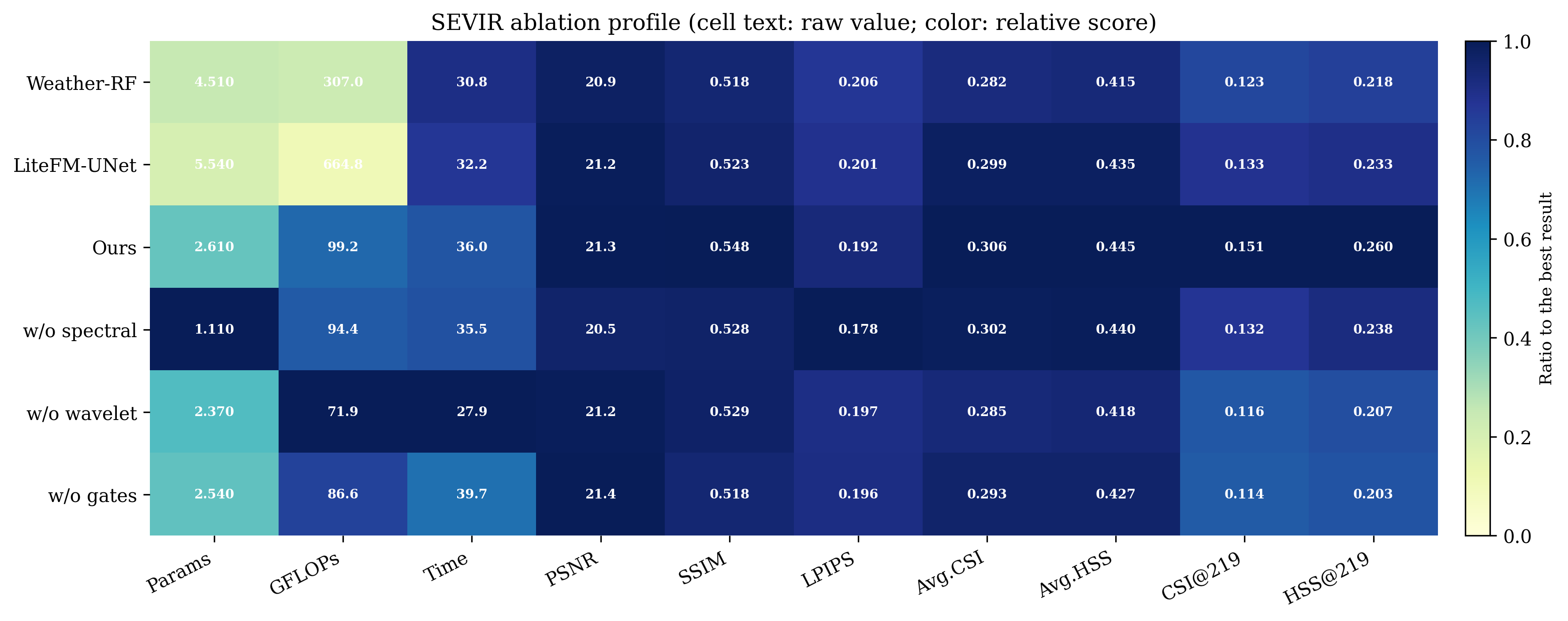}
  \caption{
  Appendix SEVIR ablation heatmap.
  Cell text gives the raw metric value, while color gives the within-column normalized score so that quality and efficiency metrics can be inspected in one panel.
  Weather-RF and Model 1 are included as reference baselines before the architectural ablations.
  }
  \label{fig:supp_ablation_heatmap}
  \vspace{-10pt}
\end{figure*}

\subsection{Diagnostic Figures}
\label{app:visual_diagnostics}

Figure~\ref{fig:supp_radar_capability} provides a normalized multi-metric capability profile for the main evaluated models.
The goal of this radar view is to summarize whether a method is balanced across reconstruction quality, high-threshold precipitation skill, and computational cost, rather than strong only on one metric.
Because the axes are normalized within each dataset and cost-oriented axes are direction-adjusted, the plot should be read as a relative diagnostic; the exact values are reported in the tables.

Figure~\ref{fig:supp_external_baseline_heatmap} expands the comparison beyond the controlled flow baseline and visualizes the broader set of CNN, transformer, generative, flow, and weather-specific baselines used in the main paper.
The heatmap is useful because these baselines differ substantially in metric scale, model size, and retrieval behavior.
The cell text retains the raw value, while the color highlights within-metric ranking, making it easier to identify whether a method's apparent strength comes from image-level averages, threshold skill, perceptual quality, or parameter efficiency.

Figure~\ref{fig:supp_threshold_curves} shows the CSI variation across precipitation thresholds instead of reporting only a single averaged event score.
This is important for satellite-to-radar retrieval because weak or non-precipitating pixels dominate the image domain, whereas intense localized echoes are sparse and more sensitive to smoothing and displacement.
The curves therefore provide a direct check of whether the model preserves skill as the event definition moves from light precipitation to heavier convective structures.

Figure~\ref{fig:supp_quality_efficiency} separates reconstruction error from sampling cost.
In pixel-space flow matching, inference requires repeated velocity-network evaluations, so quality should be interpreted together with per-sample FLOPs and parameter count.
This plot makes the quality-efficiency trade-off visible in a form that complements the main radar summary and clarifies whether reduced model size is accompanied by a meaningful change in retrieval quality.

Figure~\ref{fig:supp_ablation_heatmap} summarizes the SEVIR ablation results used to interpret the proposed spectral-local-wavelet backbone.
The heatmap compares the full model with variants that remove the spectral branch, remove the wavelet branch, or disable gated fusion, while keeping the training and sampling protocol fixed.
It is intended to connect empirical changes in quality, threshold skill, and efficiency back to the architectural components, rather than to serve as an additional baseline ranking table.

These diagnostic plots are not intended to replace the numerical tables.
They provide a compact visual check that the conclusions are not driven by a single metric and should be read together with Appendix Tables~\ref{tab:threshold_scores} and~\ref{tab:efficiency}, which provide the exact values used for comparison.

\begin{figure*}[htbp]
  \centering
  \includegraphics[width=\linewidth]{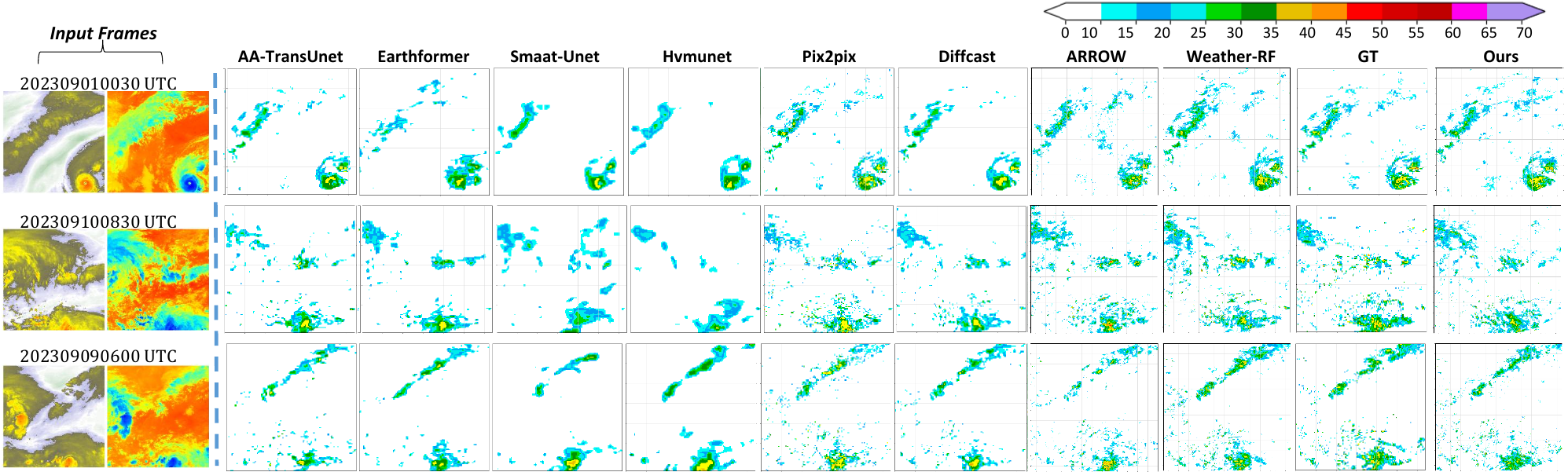}
  \caption{
  Qualitative main-experiment comparison on the Southeast China ($100.0^\circ$E--$120.0^\circ$E, $20.0^\circ$N--$40.0^\circ$N) satellite-to-reflectivity dataset.
  The panels compare satellite infrared conditions, reference reflectivity-like targets, model predictions, and absolute-error maps, providing a case-level view of displacement and intensity errors on the native regional grid.
  }
  \label{fig:china_retrieval_case1}
    \vspace{-2pt}
\end{figure*}
\subsection{Southeast China Qualitative Cases}
\label{app:china_cases}

Figure~\ref{fig:china_retrieval_case1} provides additional qualitative examples for the southeast China satellite-to-reflectivity task.
These cases are evaluated on the native $500\times500$ regional grid rather than on a resized benchmark grid, so they are intended to show whether the model preserves spatial organization at the resolution used by the operational-style retrieval setting.
The displayed panels compare the infrared satellite conditions, the reference reflectivity-like target, model predictions, and absolute-error fields.
We use these examples to inspect three failure modes that are not fully captured by average image metrics: displacement of convective cores, underestimation of locally intense precipitation, and false precipitation spread in weakly precipitating cloud regions.
The case-level evidence complements the threshold scores because high-threshold CSI and POD are sensitive to small spatial shifts in compact convective cells.

\subsection{Large-Area China Inference Case}
\label{app:large_area_case}

\begin{figure*}[htbp]
  \centering
  \setlength{\tabcolsep}{2pt}
  \begin{tabular}{cc}
    \textbf{FY-4B IR069} & \textbf{FY-4B IR107} \\
    \includegraphics[width=0.455\linewidth]{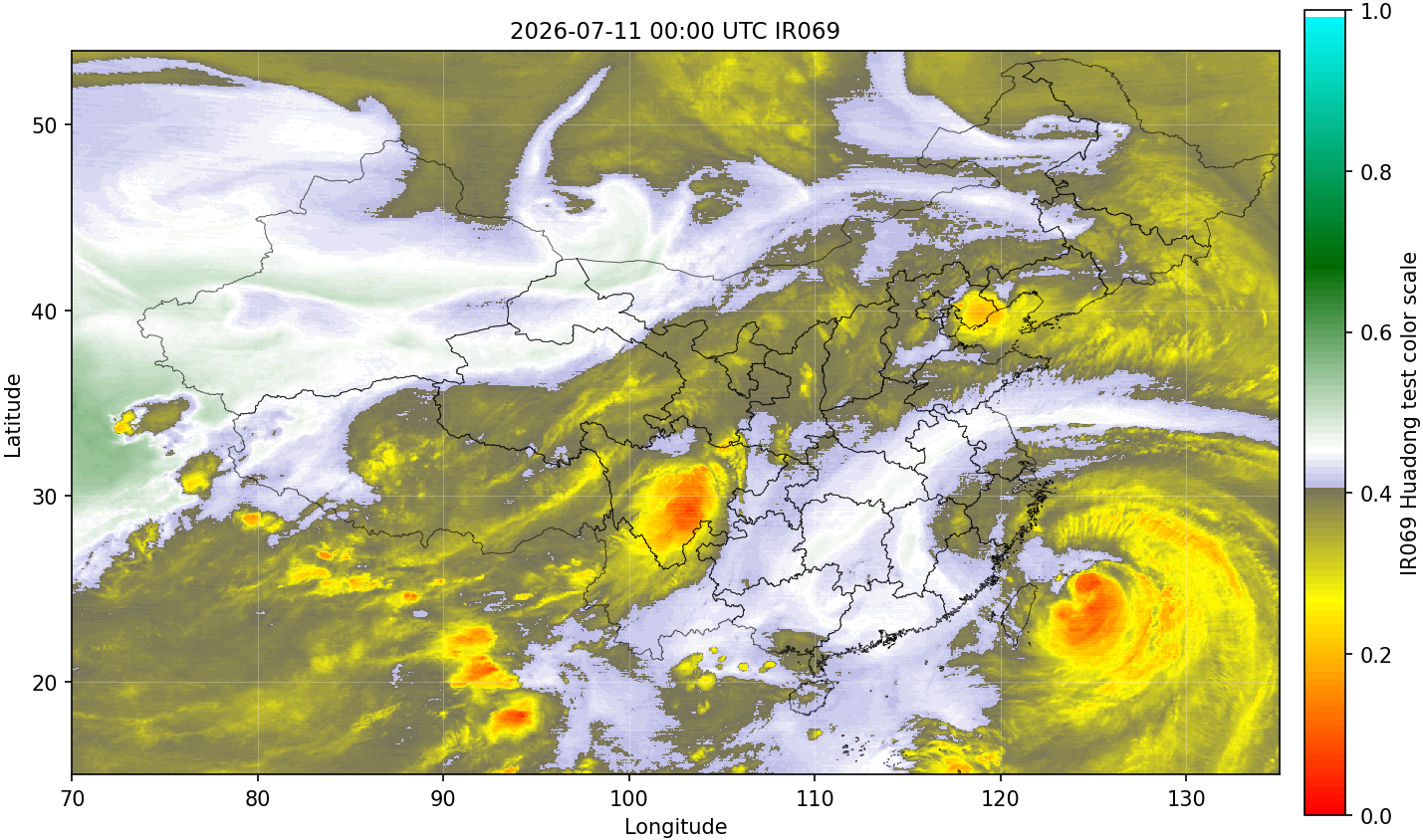} &
    \includegraphics[width=0.455\linewidth]{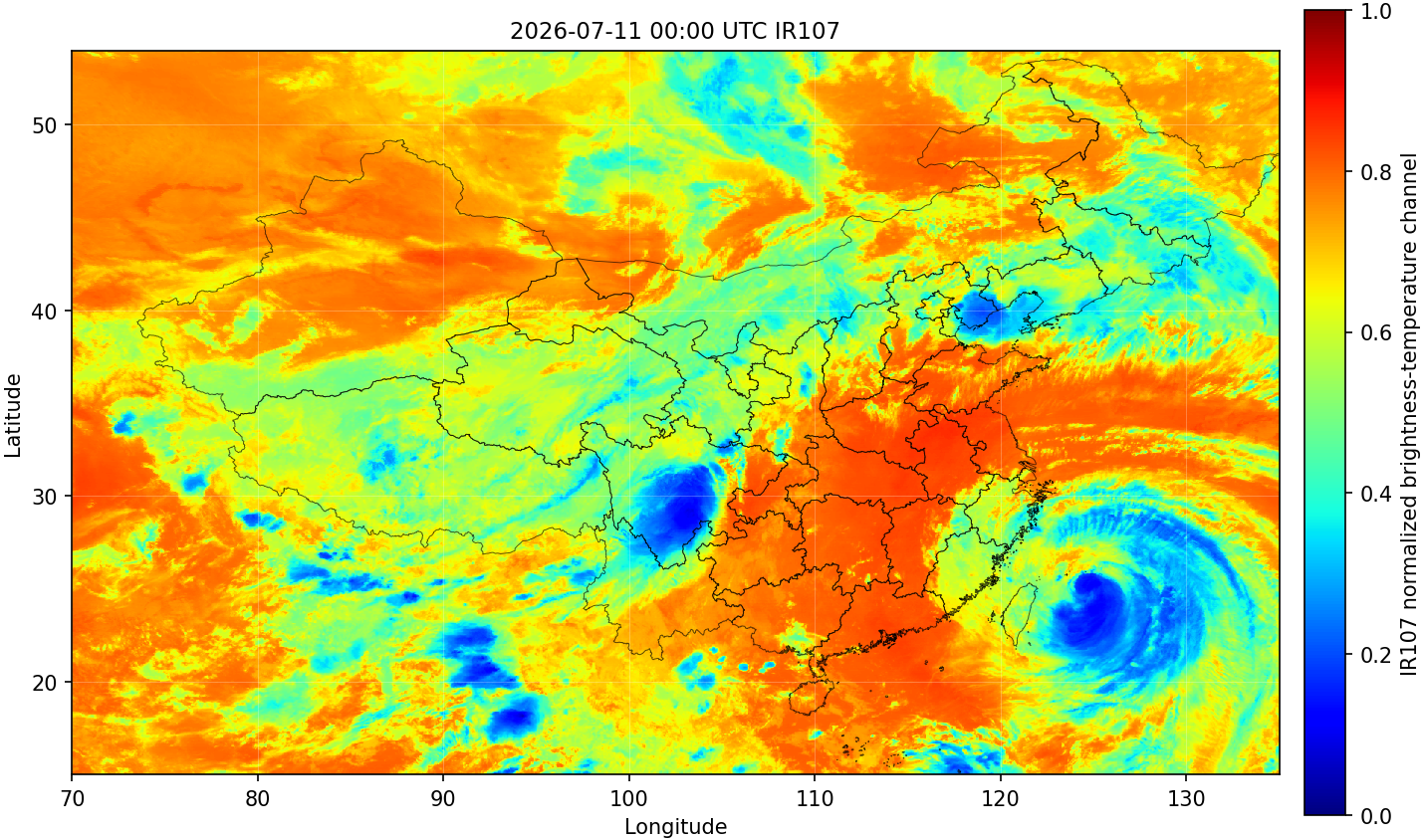} \\
    \textbf{WaveOp-LiteFM retrieval} & \textbf{GPM IMERG reference} \\
    \includegraphics[width=0.455\linewidth]{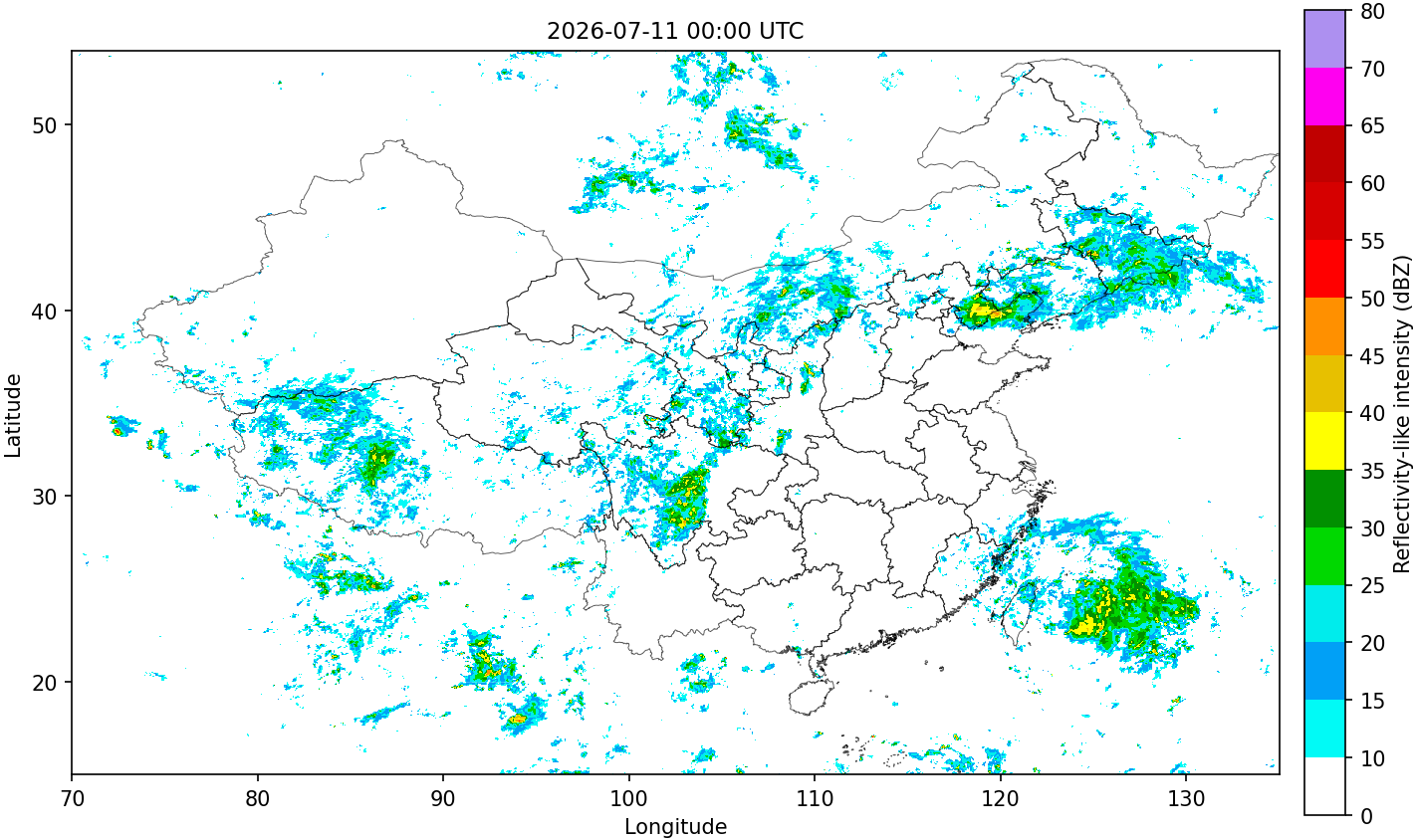} &
    \includegraphics[width=0.455\linewidth]{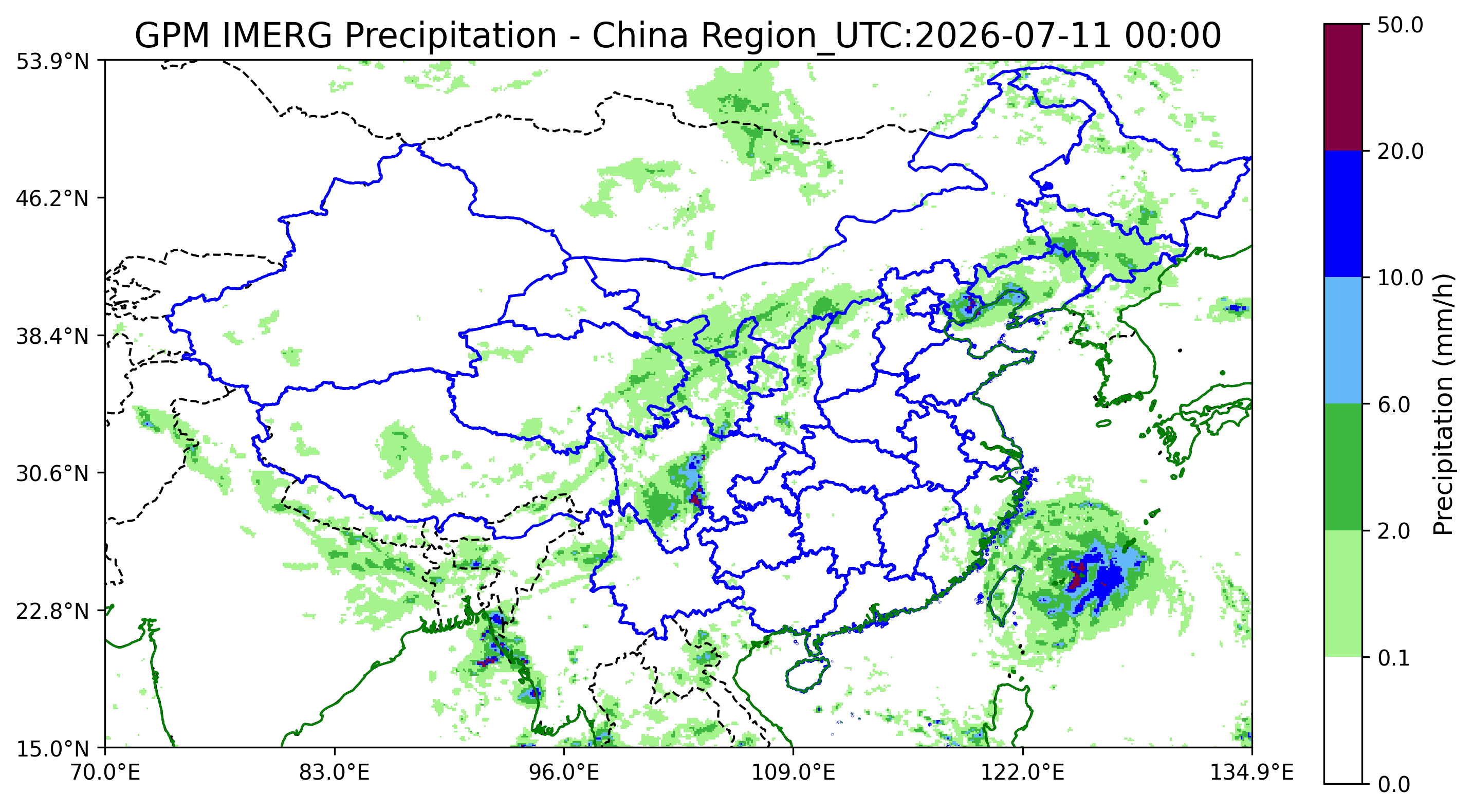}
  \end{tabular}
  \caption{
  Appendix large-area inference example for Typhoon Bavi over China at 00:00 UTC on 11 July 2026, using an FY-4B domain spanning $70.0^\circ$E--$135.0^\circ$E and $15.0^\circ$N--$54.0^\circ$N.
  The top row shows the two FY-4B infrared inputs, and the bottom row compares the stitched WaveOp-LiteFM radar-like retrieval with the corresponding GPM IMERG precipitation field at the same UTC time.
  This case is selected to illustrate the large-scale precipitation organization associated with Bavi, a recent intense typhoon case.
  Although GPM and the radar-like retrieval are different precipitation products and are not used here for pixel-wise scoring, the broadly similar precipitation organization provides an independent qualitative check that the large-area inference produces meteorologically plausible structures.
  }
  \label{fig:supp_large_area_gpm}
\end{figure*}

Figure~\ref{fig:supp_large_area_gpm} shows a stitched large-area inference example over China for Typhoon Bavi at 00:00 UTC on 11 July 2026.
The FY-4B input domain covers $70.0^\circ$E--$135.0^\circ$E and $15.0^\circ$N--$54.0^\circ$N at $0.04^\circ$ spacing, corresponding to a $975\times1625$ grid after arranging latitude from north to south.
This domain is substantially larger than the native southeast China evaluation grid and is used to test whether the fully convolutional operator backbone can be applied through tiled inference without visually obvious stitching artifacts.
The top row reports the two infrared inputs used by the southeast China model, and the bottom row compares the stitched WaveOp-LiteFM radar-like retrieval with a same-time GPM IMERG precipitation field.
Because IMERG and the radar-like retrieval are different precipitation products with different sensing physics and retrieval assumptions, this case is not used for pixel-wise scoring.
Instead, it provides an independent qualitative check that the retrieved precipitation organization remains meteorologically plausible under a broader spatial deployment setting.

\begin{table*}[t]
  \centering
  \caption{
  Threshold-based performance comparison across event intensities.
  SE China thresholds use reflectivity $Z \ge \tau$ after 0--70 dBZ
  scaling, while \sevir{} thresholds use encoded
  $\mathrm{VIL} \ge \tau$.
  }
  \label{tab:threshold_scores}

  \footnotesize
  \setlength{\tabcolsep}{3.2pt}
  \renewcommand{\arraystretch}{1.12}

  \resizebox{\textwidth}{!}{%
  \begin{tabular}{@{}l*{16}{c}@{}}
    \toprule
    \multicolumn{17}{c}{\textbf{SE China}} \\
    \midrule

    &
    \multicolumn{4}{c}{$Z \ge 10$ dBZ}
    &
    \multicolumn{4}{c}{$Z \ge 20$ dBZ}
    &
    \multicolumn{4}{c}{$Z \ge 30$ dBZ}
    &
    \multicolumn{4}{c}{$Z \ge 35$ dBZ}
    \\
    \cmidrule(lr){2-5}
    \cmidrule(lr){6-9}
    \cmidrule(lr){10-13}
    \cmidrule(lr){14-17}

    Method
    & CSI $\uparrow$ & FAR $\downarrow$ & POD $\uparrow$ & HSS $\uparrow$
    & CSI $\uparrow$ & FAR $\downarrow$ & POD $\uparrow$ & HSS $\uparrow$
    & CSI $\uparrow$ & FAR $\downarrow$ & POD $\uparrow$ & HSS $\uparrow$
    & CSI $\uparrow$ & FAR $\downarrow$ & POD $\uparrow$ & HSS $\uparrow$
    \\
    \midrule

    LiteFM-UNet
    & 0.325 & 0.380 & 0.406 & 0.417
    & 0.237 & 0.470 & 0.300 & 0.347
    & 0.147 & 0.606 & 0.191 & 0.247
    & 0.094 & 0.708 & 0.122 & 0.167
    \\

    Arrow
    & 0.313 & 0.393 & 0.392 & 0.402
    & 0.232 & 0.491 & 0.299 & 0.340
    & 0.151 & 0.633 & 0.205 & 0.252
    & 0.097 & 0.735 & 0.133 & 0.173
    \\

    Weather-RF
    & 0.367 & 0.342 & 0.453 & 0.467
    & 0.315 & 0.423 & 0.409 & 0.444
    & 0.218 & 0.557 & 0.300 & 0.347
    & 0.142 & 0.656 & 0.195 & 0.244
    \\

    \method{}
    & 0.375 & 0.330 & 0.459 & 0.477
    & 0.318 & 0.390 & 0.400 & 0.450
    & 0.224 & 0.531 & 0.301 & 0.356
    & 0.146 & 0.640 & 0.197 & 0.250
    \\

    \midrule
    \multicolumn{17}{c}{\textbf{\sevir{}}} \\
    \midrule

    &
    \multicolumn{4}{c}{$\mathrm{VIL} \ge 16$}
    &
    \multicolumn{4}{c}{$\mathrm{VIL} \ge 74$}
    &
    \multicolumn{4}{c}{$\mathrm{VIL} \ge 160$}
    &
    \multicolumn{4}{c}{$\mathrm{VIL} \ge 219$}
    \\
    \cmidrule(lr){2-5}
    \cmidrule(lr){6-9}
    \cmidrule(lr){10-13}
    \cmidrule(lr){14-17}

    Method
    & CSI $\uparrow$ & FAR $\downarrow$ & POD $\uparrow$ & HSS $\uparrow$
    & CSI $\uparrow$ & FAR $\downarrow$ & POD $\uparrow$ & HSS $\uparrow$
    & CSI $\uparrow$ & FAR $\downarrow$ & POD $\uparrow$ & HSS $\uparrow$
    & CSI $\uparrow$ & FAR $\downarrow$ & POD $\uparrow$ & HSS $\uparrow$
    \\
    \midrule

    LiteFM-UNet
    & 0.656 & 0.231 & 0.816 & 0.732
    & 0.465 & 0.360 & 0.630 & 0.591
    & 0.311 & 0.488 & 0.443 & 0.467
    & 0.133 & 0.751 & 0.222 & 0.233
    \\

    Arrow
    & 0.397 & 0.559 & 0.800 & 0.399
    & 0.349 & 0.411 & 0.461 & 0.467
    & 0.218 & 0.537 & 0.292 & 0.350
    & 0.080 & 0.853 & 0.149 & 0.146
    \\

    Weather-RF
    & 0.628 & 0.235 & 0.778 & 0.707
    & 0.445 & 0.338 & 0.576 & 0.573
    & 0.290 & 0.527 & 0.429 & 0.441
    & 0.123 & 0.778 & 0.218 & 0.218
    \\

    \method{}
    & 0.648 & 0.241 & 0.816 & 0.723
    & 0.468 & 0.344 & 0.620 & 0.595
    & 0.315 & 0.500 & 0.460 & 0.471
    & 0.151 & 0.761 & 0.289 & 0.260
    \\

    \bottomrule
  \end{tabular}%
  }

  % \vspace{-5pt}
\end{table*}
\begin{table*}[htbp]
  \centering
  \caption{
  Efficiency measurements for Model 1, Arrow, Weather-RF, and Model 4 with 20-step sampling.
  FLOPs are hook-based estimates from the implemented architecture and are reported from the current evaluation outputs.
  }
  \label{tab:efficiency}
  \setlength{\tabcolsep}{6pt}
  \renewcommand{\arraystretch}{1.08}
  \resizebox{\textwidth}{!}{
  \begin{tabular}{c l ccccccc}
    \toprule
    Dataset & Method & Input size & Params & GFLOPs/forward & GFLOPs/sample & Time (ms) $\downarrow$ & SSIM $\uparrow$ & Sampling steps \\
    \midrule
    \multirow{4}{*}{\rotatebox{90}{SE China}} & Model 1 LiteFM-UNet & $500\times500$ & 5.54M & 505.83 & 10116.57 & 416.30 & 0.555 & 20 \\
     & Arrow & $500\times500$ & 3.36M & 90.99 & 1819.70 & 160.79 & 0.549 & 20 \\
     & Weather-RF & $500\times500$ & 3.07M & 149.99 & 2999.78 & 388.27 & 0.550 & 20 \\
     & \method{} & $500\times500$ & 2.61M & 75.05 & 1500.94 & 429.22 & 0.578 & 20 \\
    \midrule
    \multirow{5}{*}{\rotatebox{90}{\sevir{}}} & Model 1 LiteFM-UNet & $128\times128$ & 5.54M & 33.24 & 664.79 & 32.25 & 0.523 & 20 \\
     & Arrow & $128\times128$ & 7.82M & 9.16 & 183.29 & 20.11 & 0.248 & 20 \\
     & Weather-RF & $128\times128$ & 4.51M & 15.35 & 307.03 & 30.83 & 0.518 & 20 \\
     & \method{} & $128\times128$ & 2.61M & 4.96 & 99.15 & 36.03 & 0.548 & 20 \\
    \bottomrule
  \end{tabular}
  }
\end{table*}

\subsection{Detailed Threshold and Efficiency Results}
\label{app:detailed_results}

The main paper uses the radar summary plot to avoid duplicating large numerical tables.
For completeness, Appendix Table~\ref{tab:threshold_scores} reports the full threshold contingency scores, and Appendix Table~\ref{tab:efficiency} reports the detailed efficiency measurements used to construct the main radar view.
The main comparison table in the paper should be read together with these Appendix tables.
On southeast China, the 20-step sampling cost changes from 10116.6 GFLOPs/sample for LiteFM-UNet to 1500.9 for \method{}, while the trainable parameters change from 5.54M to 2.61M.
On \sevir{}, the corresponding cost changes from 664.8 to 99.2 GFLOPs/sample.
For the heavy-event rows in the main table, the southeast China threshold is 35 dBZ and the \sevir{} threshold is encoded VIL@219.
These values are not repeated in the main text so that the main experiment narrative can focus on the scientific interpretation: whether the lightweight operator backbone preserves sparse intense precipitation skill while reducing repeated velocity-network computation.

% \FloatBarrier

\subsection{Ablation Design}
\label{app:ablation_design}

The ablation study isolates three architectural choices in WaveOp-LiteFM.
The first ablation removes the spectral branch, testing whether low-frequency operator mixing is necessary under the \sevir{} setting.
The second removes the wavelet threshold branch, testing the contribution of explicit high-frequency shrinkage.
The third disables gated fusion, testing whether input-adaptive branch and skip mixing improves the quality-efficiency trade-off.
All ablation models retain the same optimizer, data split, flow objective, and sampler as the full model.
Appendix Figure~\ref{fig:supp_ablation_heatmap} summarizes these variants across quality, threshold skill, and efficiency metrics, providing a compact diagnostic view of how each architectural component affects the final retrieval behavior.

\subsection{Tiled Large-Area Inference Protocol}
\label{app:tiled_inference}

WaveOp-LiteFM is fully convolutional apart from Fourier and Haar operations that can be evaluated on arbitrary spatial grids within memory limits.
For regions larger than the training crop or GPU memory budget, large-area retrieval is performed by overlapping tiled inference followed by stitching.
The 2026 FY-4B large-area input covers $70.0^\circ$E--$135.0^\circ$E and $15.0^\circ$N--$54.0^\circ$N at $0.04^\circ$ spacing, giving a $975\times1625$ grid after the north-to-south latitude ordering is applied.
The two normalized infrared channels are denoted by
$c\in[0,1]^{2\times H\times W}$.
We use square patches of size $P=512$ with overlap $O=128$, so the stride is $S=P-O=384$.
Let $\mathcal{I}$ be the set of upper-left patch coordinates $(i,j)$ generated by this stride rule, with an additional terminal patch added when needed so that the full grid is covered.

For each patch coordinate $(i,j)$, the satellite condition and initial noise are cropped from the same full-domain tensors,
\begin{align}
  c_{ij}=\mathcal{C}_{ij}(c), \qquad
  z_{ij}=\mathcal{C}_{ij}(z),
\end{align}
where $\mathcal{C}_{ij}(\cdot)$ extracts the $P\times P$ window whose upper-left coordinate is $(i,j)$.
The noise tensor $z\sim\mathcal{N}(0,I)$ is sampled once over the padded full domain.
Using a single global noise realization avoids changing the stochastic initial condition independently from tile to tile.
Each patch is sampled with the same Euler flow-matching solver used in the main experiments,
\begin{equation}
  x^{(k+1)}_{ij}
  =
  x^{(k)}_{ij}
  +
  \Delta t\,
  v_\theta\!\left(x^{(k)}_{ij},t_k,c_{ij}\right),
  \qquad
  x^{(0)}_{ij}=z_{ij},
\end{equation}
for $K=20$ steps in the default full run.
The resulting patch prediction is denoted by $\hat{y}_{ij}\in[0,1]^{1\times P\times P}$.

To suppress tile-edge discontinuities, overlapping predictions are blended with a separable Hann window.
Let $h\in\mathbb{R}^{P}$ be the one-dimensional Hann window clipped by a small floor, and define
\begin{equation}
  w(u,v)=\max(h_u,\epsilon)\max(h_v,\epsilon),
  \qquad \epsilon=0.05 .
\end{equation}
The stitched normalized retrieval is computed by weighted accumulation.
For each full-domain pixel $(p,q)$, we accumulate
\begin{align}
  A(p,q)
  &=
  \sum_{(i,j)\in\mathcal{I}}
  \mathbf{1}_{(p,q)\in\Omega_{ij}}\,
  w_{ij}(p,q)\,
  \hat{y}_{ij}(p-i,q-j), \\
  B(p,q)
  &=
  \sum_{(i,j)\in\mathcal{I}}
  \mathbf{1}_{(p,q)\in\Omega_{ij}}\,
  w_{ij}(p,q),
\end{align}
where $w_{ij}(p,q)=w(p-i,q-j)$.
The final stitched field is
\begin{equation}
  \hat{y}(p,q)=A(p,q)/B(p,q),
\end{equation}
where $\Omega_{ij}$ is the spatial support of the patch.
After stitching, the padded margins are removed and the output is clipped to $[0,1]$.
For visualization and threshold interpretation, the normalized field is mapped to a reflectivity-like scale by
\begin{equation}
  \hat{Z}(p,q)=70\,\hat{y}(p,q) .
\end{equation}

This procedure is a deployment mechanism rather than a claim of distribution-free generalization.
When the target region, season, sensor geometry, or precipitation regime changes, performance should be revalidated with the same threshold and image metrics used in the main experiments.
Figure~\ref{fig:supp_large_area_gpm} provides the corresponding large-area qualitative example, pairing the two FY-4B inputs and the stitched retrieval with a same-time GPM IMERG reference field.

\end{document}